\documentclass[journal]{IEEEtran} 
\pdfoutput=1
\usepackage{amsthm}  
\usepackage{mathrsfs}
\usepackage{epsfig}
\usepackage{amsmath}
\usepackage{cite}
\usepackage{amsfonts}
\usepackage{amssymb}
\usepackage{mathpazo}
\usepackage{enumerate}
\usepackage{times}
\usepackage{graphicx}
\usepackage{epstopdf}

\usepackage[monochrome]{color}  
\usepackage{cases}

\usepackage{algorithm}
\usepackage{algpseudocode}

\newtheorem{theorem}{Theorem}
\newtheorem{corollary}{Corollary}
\newtheorem{definition}{Definition}
\newtheorem{lemma}{Lemma}

\newtheorem{assumption}{Assumption}
\newtheorem{remark}{Remark}
\newtheorem{example}{Example}

\begin{document}

	\title{On Synchronization of Dynamical Systems over    Directed Switching Topologies: An Algebraic\\ and Geometric {\color{blue}Perspective}}

	\author{Jiahu Qin, \IEEEmembership{Senior Member,~IEEE},~Qichao Ma,%
		~Xinghuo Yu, \IEEEmembership{Fellow,~IEEE},~and Long Wang, \IEEEmembership{Member,~IEEE}
		\thanks{%
			J.~Qin and Q.~Ma are with the Department of Automation, University of Science and Technology of China,
			Hefei 230027, China
			(e-mail: \texttt{jhqin@ustc.edu.cn}; \texttt{mqc0214@mail.ustc.edu.cn}).}
	
		\thanks{X.~Yu is with the School of Engineering, RMIT University,
			Melbourne, VIC 3001, Australia
			(e-mail: \texttt{x.yu@rmit.edu.au}).}
		
		\thanks{L.~Wang is with the Center for systems and Control, College of Engineering, Peking University, Beijing 100871, China
			(e-mail: \texttt{longwang@pku.edu.cn}).}
	}
	
	\maketitle

	
\begin{abstract}
In this paper, we aim to investigate the synchronization problem of dynamical systems, which can be of generic linear or Lipschitz nonlinear type, communicating over directed switching network topologies. {\color{blue}A mild} connectivity assumption on the switching topologies is imposed, which allows them to be directed and jointly connected.	We propose a novel analysis framework from both algebraic and geometric perspectives to justify the attractiveness of the synchronization manifold. Specifically, it is proven that the complementary space of the synchronization manifold can be spanned by certain subspaces. These subspaces can be the eigenspaces of the nonzero eigenvalues of Laplacian matrices in linear case. They can also be
subspaces in which the projection of the nonlinear self-dynamics still retains the Lipschitz property. This allows to  project the states of the dynamical systems into these subspaces and transform the synchronization problem under consideration equivalently into a convergence one of the projected states in each subspace.  Then, assuming the joint connectivity condition on the communication topologies, we are able to work out a simple yet effective and unified convergence analysis for both types of dynamical systems. More specifically, for partial-state coupled generic linear systems, it is proven that synchronization can be reached if an extra condition, which is easy to verify in several cases, on the system dynamics is satisfied. For Lipschitz-type nonlinear systems with positive-definite inner coupling matrix, synchronization is realized if the coupling strength is strong enough to stabilize the evolution of the projected states in each subspace under certain conditions. The above claims generalize the existing results concerning both types of dynamical systems to so far the most general framework. Some illustrative examples are provided to verify our theoretical findings.
\end{abstract}
	
	\begin{IEEEkeywords}
	Synchronization control, directed switching topology, linear generic system,  Lipschitz-type nonlinear system.
	\end{IEEEkeywords}

\section{Introduction\label{sec:intro}}

Over the last couple of years, consensus and synchronization problems have been popular subjects in systems and control \cite{Wieland2011,Ren2005,QinTCYB2016}, inspired by their applications in physics, social sciences, biology, and engineering \cite{Liu2012,Passino2002,Book-BiologicalTime}. The essence of these kinds of problems is
the collective objective to reach agreement about some variables of interest 
\cite{Qin2015TNNLS,Qin2013AUCC,QinTAC2016,QinTCYB2016}. A widely used control protocol to achieve the above-mentioned goal is the linear controller using nearest neighbors' information \cite{Jadbabaie2003}.
In determining the collective behavior using the distributed linear controller, three different factors are fundamental, namely, the self-dynamics, the coupling configuration (e.g., type and strength of couplings), and the coupling topology \cite{Jalili2013,Wieland2011}.  To date, intensive analyses on such issues as how these factors influence the collective behaviors of networked systems has been conducted and fruitful conclusions have been obtained. Synchronization over switching communication topology is one of these issues and is attracting great attentions. {\color{blue}So far, necessary and/or sufficient connectivity conditions to achieve consensus for first-order dynamics have been very well developed, e.g., \cite{CaoMSIAM2008,Ren2005,Shi2013,AndersonTAC2017,XiaoTAC2028,ChenTAC2017,TouriTAC2011}.}  Unfortunately, when it comes to higher-order linear systems or nonlinear systems with complex self-dynamics and coupling configuration, there are still many open problems. We aim to further address synchronization of dynamical systems over switching topology driven by static controller in this work.

For inter-connected generic linear systems and nonlinear systems, (common/multiple) Lyapunov method is commonly adopted to perform synchronization
analysis \cite{KimAUTO2013,ChenAUTO2016}. The Lyapunov function is appropriately designed such that the factors as coupling topology are involved
\cite{WenTNNLS2015}. The difficulty of applying Lyapunov method is each switched sub-system may not be a convergent one (because the Laplacian matrix has
multiple  zero eigenvalues). To overcome this difficulty, it is usually required that the communication graph has a well connectivity property
\cite{KimAUTO2013,BackTAC2017,WenTNNLS2015}. Specifically, in \cite{KimAUTO2013} synchronization among partial-state coupled  identical linear systems (viz., input matrix is not invertible)
 using dynamic controller is investigated, where the communication topology is assumed to have a
well-defined average that is connected.
In \cite{WenTNNLS2015}, the authors seek synchronization via multiple Lyapunov function approach
assuming that the communication graph is frequently connected, that is, the graph is connected over at least one sub-interval for a period of time.
Matrix inequalities are proposed with respect to system matrix and Laplacian matrix to guarantee practical synchronization
\cite{WangAUTO2015} in the presence of input disturbance, where the communication graph remains connected all the time. Very recently, a dynamic controller is designed in \cite{BackTAC2017} to address the bounded synchronization for uncertain linear
systems which communicate over frequently connected undirected graphs.

The contraction analysis \cite{ChenAUTO2016,YangTCSII2016} is another approach to dealing with synchronization problem over switching topology. This approach focuses on deriving the contraction property of synchronization error/disagreement vector. In particular, it is aimed to show that the synchronization error decreases strictly over sufficiently long time, usually with mild connectivity condition  at the cost of additional constraints on system model \cite{YangTCSII2016} or stringent sufficient algebraic condition \cite{ChenAUTO2016}.
Specifically, ref. \cite{YangTCSII2016} considers the synchronization problem among nonlinear system dynamics, which satisfies Lipschitz condition.
By considering the contraction of the norm of the state deviation, the factors related to nonlinear system dynamics, switching communication graph that is jointly strongly connected, and coupling strength are implicitly involved in the sufficient condition taking algebraic form. In ref.~\cite{ChenAUTO2016}, similar technique to that used in \cite{YangTCSII2016} is applied to derive synchronization condition  with relaxed joint connectivity condition. However, from the sufficient algebraic condition provided therein, one may not figure out how the switching scheme, the coupling topology as well as the coupling strength influence the synchronization behavior, which is one of the most fundamental issues in examining the collective behavior of dynamical systems.   Although ref. \cite{QinTAC2014} succeeds in addressing  this issue by working on the networks of linear systems under mild constraint on communication topology, it is, however, assumed that the input matrix is invertible.



In this paper,  we propose a novel analysis framework, which is totally different from most existing works, {\color{blue}from an unified algebraic and
geometric perspective to revisit the synchronization problems for both generic linear systems and Lipschitz-type nonlinear systems over switching directed
topologies. It is interesting to observe that the complementary space of the synchronization manifold can be spanned by certain subspaces. These
subspaces can be the eigenspaces of the nonzero eigenvalues of Laplacian matrices induced from communication topologies in linear case. They can also be
subspaces in which the projection of the nonlinear self-dynamics still retains the Lipschitz property.} This allows us to project the states of the
systems onto these subspaces.  Subsequently, to guarantee synchronization, it suffices to show that the states of the systems vanish along each of these
subspaces by employing techniques developed from matrix analysis and stability theory. {\color{blue}To the best of our knowledge, no approach has been developed so far that can simultaneously deal with the synchronization analysis of both generic linear and Lipschitz-type nonlinear systems.}

{\color{blue}We are able to, with the above analysis framework, tackle the difficulty confronted by the approach in \cite{NiSCL2010,ValcherAUTO2017} and \cite{SuTAC2012} that the associated Laplacian matrices may have eigenvalue zero with algebraic multiplicity larger than
one and cannot apply Lyapunov function method to the cases with directed and disconnected topologies.}   Moreover, different from \cite{ChenAUTO2016} and \cite{WenTNNLS2015} where the contraction analysis is performed directly with respect to state deviation, we analyze the contraction in the subspaces in which the projection of nonlinear system dynamics still retains Lipschitz property. Hence, the contraction property, which is guaranteed in \cite{ChenAUTO2016,WenTNNLS2015} by an algebraic condition,  follows easily with strong couplings.

With these observations, we are able to work out the following contributions under {\color{blue}a mild} joint connectivity condition.

\begin{enumerate}
\item It is proven that synchronization for linear partial-state coupled systems can be achieved if an algebraic  condition with regard to system dynamics and Laplacian matrix is satisfied. This generalizes the results in \cite{NiSCL2010} and \cite{SuTAC2012} from undirected communication topologies to the directed case.   As a byproduct of the observation that eigenvalue zero of the graph Laplacian matrix is semisimple, we show that if the communication graph switches slowly, then synchronization can be achieved for a class of marginally stable and positive linear systems under the joint connectivity condition. A lower bound of the dwell time is also explicitly specified to guarantee the synchronization.

\item For  Lipschitz-type nonlinear systems, it is found that with sufficiently strong couplings to ensure the decay of the projected states onto the subspaces in which the Lipschitz property of nonlinear system still holds, the synchronization can be guaranteed. This reveals that the desynchronization coming from self-dynamics should be dominated by the synchronization contributed by the jointly connected communication graph provided a certain geometric property of the subspaces holds.  Although sufficient conditions that are required to synchronize the  Lipschitz-type nonlinear  systems have also been developed, e.g., in refs. \cite{ChenAUTO2016} and \cite{WenTNNLS2015}, no such information as how the self-dynamics, the coupling strength and/or coupling topologies influence the synchronization behavior have been revealed intuitively and explicitly.

\end{enumerate}

The remainder of the paper is arranged as follows. In Section \ref{sec:preliminary}, we introduce the relevant graph notions and formulate the problem. Some technical lemmas are provided in Section III. The evolution analysis of the projection state in a fixed interval is provided in Section IV, followed by Section V where we provide the main results on synchronization by invoking the joint connectivity condition. Some technical analyses of the main results are presented in Section VI. The paper is concluded at last in Section \ref{sec:conclusion}

Notations:\ Let $\|x\|$ denote the Euclidean norm of a
finite dimensional vector $x$. Denote by $I_{n}$ the identity matrix (if the subscript is dropped, $I$ denotes the identity matrix of compatible dimension)
and by $0_{n\times n}$ the zero matrix in $\mathbb{R}^{n\times n}$.
  Let $\mathrm{diag}\{a_{1},\ldots,a_{q}\}$
denote the diagonal matrix with $a_{i}$ being the $i$-th diagonal
element. Let $\mathrm{Ker}(L)$ and $\mathrm{Ran}(L)$ denote the kernel and  range space of a square matrix $L$, respectively. 

\section{Graph Theory and Problem Formulation\label{sec:preliminary}}
\subsection{Graph and Matrix Theory Notions}
The interaction topology of a collection of systems is represented by the directed graph {\color{blue}$\mathcal{G}(t)=(\mathcal{V},\mathcal{E}(t),  \mathcal{W}(t))$} of order $N$ with a finite nonempty set of nodes $\mathcal{%
V=}\left\{ 1,2,\ldots ,N\right\} ,$ a set of edges
${\color{blue}\mathcal{E}(t)}
\mathcal{\subset V\times V},$ and a weighted \textit{adjacency matrix} ${\color{blue}\mathcal{W}(t)}=%
\left[ a_{ij}(t)\right] \in \mathbb{R}^{N\times N}$, where $a_{ij}(t)$ is
the weight, also called coupling strength in this work, of the
directed edge $(j,i)$ satisfying $a_{ij}(t)>
0$ if $(j,i)$ is an edge of {\color{blue}$\mathcal{G}(t)$} and $a_{ij}(t)=0$ otherwise. 
Moreover, we assume $a_{ii}(t)\equiv0$ for all $i\in \mathcal{V}.$
The Laplacian matrix {\color{blue}$L(\mathcal{G}(t))$} of
${\color{blue}\mathcal{G}(t)}=(\mathcal{V},{\color{blue}\mathcal{E}(t),\mathcal{W}(t)})$ is defined as
$L(t)=\mathrm{diag}\{\Delta_{1}(t),\ldots,\Delta_{N}(t)\}-{\color{blue}\mathcal{W}(t)}$, where
$\Delta_{i}(t)=\sum\nolimits_{j=1}^{N}a_{ij}(t),$ $i=1,\ldots,N$
\cite{AlgebraicGT}. An important fact of $L(t)$ is that
$\mathbf{1}_{N}$ is a right eigenvector associated with
eigenvalue $0$ \cite{AlgebraicGT}.
A \textit{directed path} is a sequence of edges in a directed graph
of the form $(i_{1,}i_{2}),$ $(i_{2,}i_{3}),\ldots,
(i_{q-1},i_{q}).$
A digraph \textit{has a directed
spanning tree} if there exists at least one node, called the root,
having a directed path to every other node.

For simplicity, let $\{\mathcal{G}_k |k\in \mathcal{P}\}$ denote the set of all
possible interaction graphs, each associated with the Laplacian
$L_k$ for $k\in\mathcal{P}$. Herein $\mathcal{P}=\{1,\ldots,Q \}$ with $Q>1$ being an integer. Consider an infinite sequence of nonempty,
bounded, and contiguous time intervals $[t_k,t_{k+1}),k=0,1,\cdots,$ with $t_0=0$ and $t_{k+1}-t_k\leq T_{\max}$ where $T_{\max}>0$.
In each interval $[t_k, t_{k+1})$, there is a sequence of non-overlapping
subintervals $[t_{k_0},t_{k_1})$, $[t_{k_1},t_{k_2})$, $\cdots$, $[t_{k_{m_k-1}},t_{k_{m_k}})$ with $t_{k_0}=t_k$, $t_{k_{m_k}}=t_{k+1}$ satisfying $t_{k_{j+1}}-t_{k_j} \geq T_{\min}$, $0\leq j \leq m_k-1,$ for {\color{blue}an
integer $m_k\geq 1$ and a positive constant $T_{\min}$ which is also coined as the dwell time in the literature}. The digraph $\mathcal{G}(t)$ remains unchanged during each subinterval $[t_{k_l},t_{k_{l+1}})$ and switches at $t_{k_{l+1}}$. In particular, define a right continuous switching signal $\sigma(t):[0,+\infty)\to\{1,\ldots,Q \}$ and the dynamically changing
digraph is denoted by ${\color{blue}\mathcal{G}_{\sigma(t)}}=(\mathcal{V},{\color{blue}\mathcal{E}_{\sigma(t)}},[a^{\sigma(t)}_{ij}])$ (with Laplacian matrix $L_{\sigma(t)}$).

\begin{definition}[Union of Graphs\cite{Ren2005}]
The union of a collection of graphs $\{{\color{blue}\mathcal{G}_i}\}_{i=1}^p$, each of order $N$, is a graph with node set given by $\mathcal{V}({\color{blue}\mathcal{G}_i}),\forall i$ and edge set given by $\cup_{i=1}^p {\color{blue}\mathcal{E}(\mathcal{G}_i)}$. The union graph across any time interval $[t_k, t_{k+1})$ is defined by ${\color{blue}\mathcal{G}_{\mathrm{uni}}}=\cup_{t\in[t_k, t_{k+1})}{\color{blue}\mathcal{G}_{\sigma(t)}}$.
\end{definition}

\begin{definition}[Generalized Eigenvector \cite{HornMA2012}]
If $A$ is an $n\times n$ matrix, a generalized eigenvector of $A$ corresponding to the eigenvalue $\lambda$ is a nonzero vector $v$ satisfying
$
(A-\lambda I)^p v=0,
$
for some positive integer $p$. Equivalently, it is a nonzero element of the nullspace of $(A-\lambda I)^p$. Specifically, if $p=1$, then the generalized eigenvector becomes the eigenvector.
\end{definition}

\subsection{System Model and Problem of Interest\label{subsec:problem formulation}}
Consider the following {\color{blue}partial-state} coupled linear systems
\begin{align}\label{linear-system-dynamics}
\dot{x}_i=Ax_i+\phi BK\sum_{j=1}^N a_{ij}^{\sigma(t)}(x_j-x_i),
\end{align}
while the inter-connected nonlinear systems are described by
\begin{align}\label{nonlinear-system-dynamics}
\dot{x}_i=f(x_i)+\phi\sum_{j=1}^N a_{ij}^{\sigma(t)}\Gamma(x_j-x_i),
\end{align}
for $i=1,\ldots,N$, where $x_i$ denotes the state of the $i$-th agent, $A\in\mathbb{R}^{n\times n},B\in\mathbb{R}^{n\times m},K\in\mathbb{R}^{m\times n}$ is the feedback matrix to be designed, $f:\mathbb{R}^n\to \mathbb{R}^n$ is a continuous function,  $\Gamma$ is a positive diagonal matrix, and $\phi>0$ is the coupling strength.

Our objective in this paper is to analyze under what kind of conditions synchronization can be achieved  for \eqref{linear-system-dynamics} and \eqref{nonlinear-system-dynamics} under the \textit{joint connectivity  condition}, which is stated as follows.

\begin{assumption}\label{connectivity-assumption}
There exists a positive constant $T$ such that the union graph across any time interval with length $T$ contains a directed spanning tree, i.e., $\cup_{t\in [t_0,t_0+T)}{\color{blue}\mathcal{G}_{\sigma(t)}}$ contains a directed spanning tree for any $t_0\geq 0$.
\end{assumption}

 In view of Assumption \ref{connectivity-assumption},  assume throughout this paper, without loss of generality, that the union of communication graphs over $[t_k,t_{k+1})$ contains a directed spanning tree, that is, $\cup_{t\in[t_k,t_{k+1})}{\color{blue}\mathcal{G}_{\sigma(t)}}$ contains a directed spanning tree.
Besides, our results to be established also base on the following technical assumptions.
{\color{blue}\begin{assumption}\label{linear-assumtion}
The matrix pair $(A,B)$ is stabilizable, i.e., there exists a compatible matrix $K$ such that $A-BK$ is Hurwitz.
\end{assumption}}
\begin{assumption}\label{Lipschitz-assumtion}
	The nonlinear function $f(\cdot)$ satisfies Lipschitz condition with Lipschitz constant being $\rho>0$, i.e., $\|f(x)-f(y)\|\leq \rho\|x-y\|$, $\forall x,y\in\mathbb{R}^n$.
\end{assumption}
\begin{remark}
{\color{blue}Assumption \ref{connectivity-assumption} imposes a joint connectivity assumption on switching communication graph, which is milder than those considered in existing works for synchronization of linear or Lipschitz-type nonlinear systems \cite{WenTNNLS2015,ValcherAUTO2017}. It is worth pointing out that Assumption \ref{connectivity-assumption} is not the weakest connectivity condition. Weaker constraints on connectivity include infinite joint-connectivity \cite{TouriTAC2011} and extensible joint-connectivity \cite{ChenTAC2017}.  Assumption \ref{linear-assumtion} is a necessary condition for consensusability  of linear systems via state-feedback controller \cite{MaTAC2010}, while Assumption \ref{Lipschitz-assumtion} is satisfied by many well-known systems, such as Lorenz systems \cite{WenTNNLS2015}.}
\end{remark}
\section{{\color{blue}Technical Lemmas}\label{sec:Main Results}}
In this section, we shall present several useful results. They lay the foundation of the analysis framework to be developed and enlighten the proof of our main results. Please refer to the Appendix for the proofs of all the lemmas proposed in this paper.

The following is a result on spectral property of non-symmetric
Laplacian matrix which has been proved earlier by \cite{Caughman}\footnote{The result on symmetric Laplacian matrix can be found in \cite{Jadbabaie2003}.}.

{\color{blue}\begin{lemma}\label{semi-simple-zero-eigenvalue}
Given any Laplacian matrix of a non-negatively weighted graph, its zero eigenvalue is semi-simple, that is, the algebraic multiplicity of the zero eigenvalue equals to the geometric multiplicity.
\end{lemma}}

\begin{lemma}\label{kernel-intersect}
	Given a collection of non-negatively weighted graphs $\{{\color{blue}\mathcal{G}_i}\}_{i=1}^p$, each of order $N$, if $\cup_{j=1}^p {\color{blue}\mathcal{G}_j}$ contains a directed graph, then 1) $\cap_{j=1}^p \mathrm{Ker}(L_j)=\mathrm{span}\{\mathbf{1}_N\}$ and 2) $\mathrm{span}\{\mathrm{Ran}(L_1)\cup\cdots\cup \mathrm{Ran}(L_p)\}\oplus \mathrm{span}\left\{\mathbf{1}_N\right\}=\mathbb{R}^N$, where $L_j$ is the Laplacian matrix of ${\color{blue}\mathcal{G}_j}$ for $j=1,\ldots,p$.
\end{lemma}

\begin{example}[An Illustrative Example of Lemmas \ref{semi-simple-zero-eigenvalue} and \ref{kernel-intersect}]
	\begin{figure}[h]
		\centering\includegraphics[width=7cm]{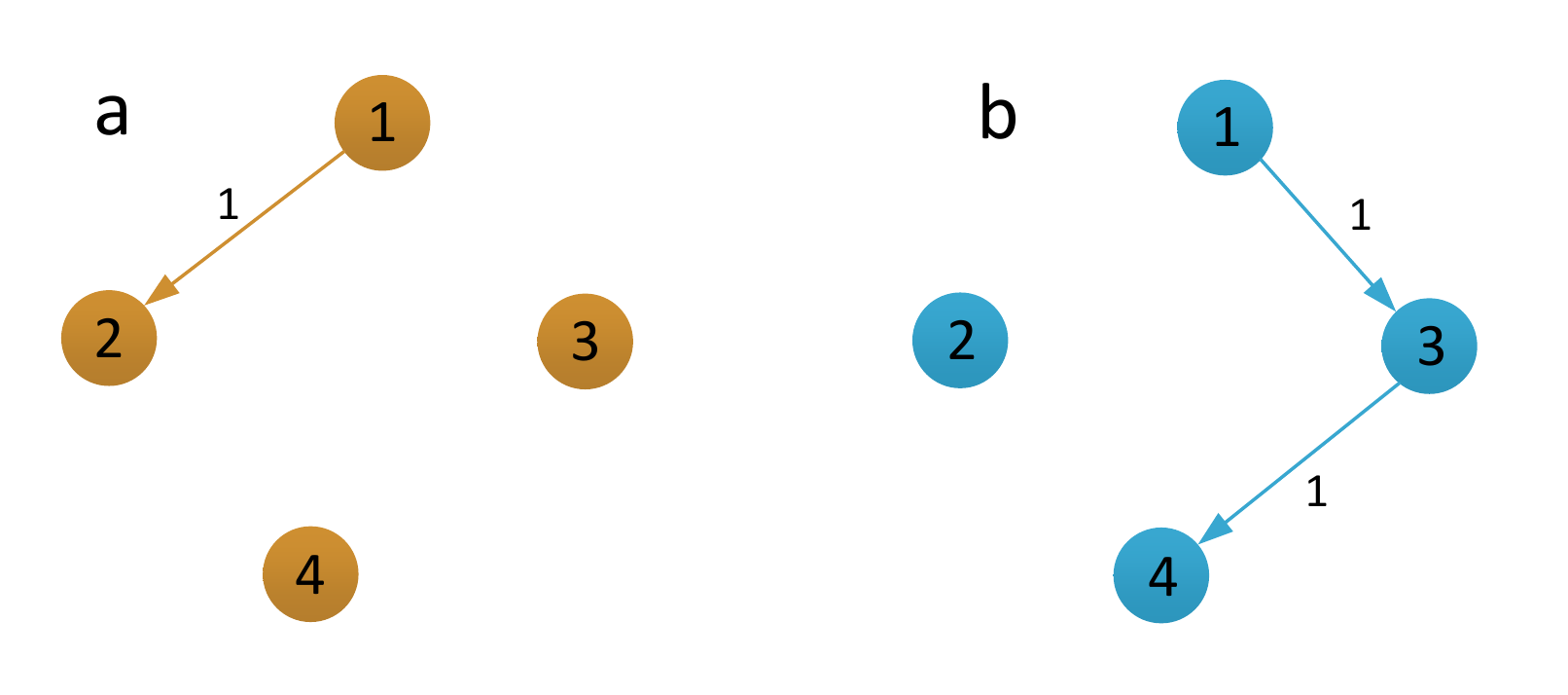}\caption{Two illustrative graphs $G_a$ and $G_b$. The union of $G_a$ and $G_b$ contains a directed spanning tree.}\label{illustrative-graph}
		\centering\includegraphics[width=7cm]{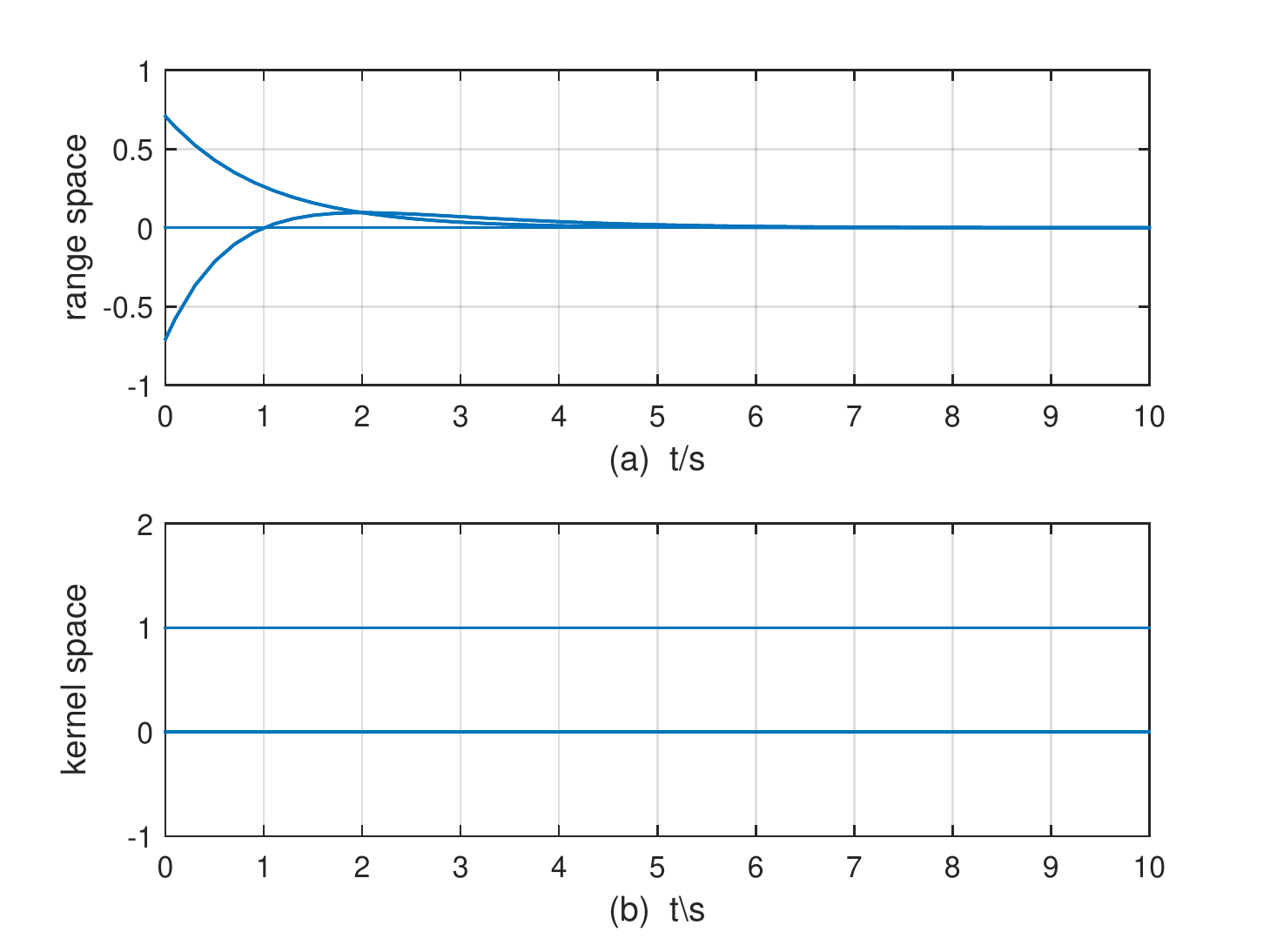}\caption {Sub-figure (a) depicts the evolution of the system states when the initial value is chosen to be $[0,0,0.7071,-0.7071]^{\mathrm{T}}$, which lies in the range space of $L_b$. Sub-figure (b) shows the evolution of the system states when the initial value is chosen to be $[0,1,0,1]$, which is contained in the kernel space of $L_b$.}\label{Evolution-Exp_1}
	\end{figure}
	Consider two Laplacian matrices $L_a$ and $L_b$ which  are respectively defined as
	\begin{align*}
	L_a=\begin{bmatrix}0&0&0&0\\-1&1&0&0\\0&0&0&0\\0&0&0&0\end{bmatrix},\;\;L_b=\begin{bmatrix}0&0&0&0\\0&0&0&0\\-1&0&1&0\\0&0&-1&1\end{bmatrix}.
	\end{align*}
It is easy to know that $\mathrm{Ker}(L_a)$ is spanned by the vectors $v_1=[0,0,1,0]^{\mathrm{T}}$, $v_2=[0, 0, 0, 1]^{\mathrm{T}}$, and $v_3=[0.7071,0.7071,0,0]^{\mathrm{T}}$. While $\mathrm{Ker}(L_b)$  is spanned by $u_1=[0,1,0,0]^{\mathrm{T}}$ and $u_2=[0.5774,0,0.5774,0.5774]^{\mathrm{T}}$. Moreover $\mathrm{Ran}(L_a)$ is spanned by $[0,1,0,0]^{\mathrm{T}}$, and $\mathrm{Ran}(L_b)$ is spanned by $[0,0,1,0]^{\mathrm{T}}$ and $[0,0,0,1]^{\mathrm{T}}$. It is easily verified that $\mathrm{Ker}(L_a)\cap \mathrm{Ker}(L_b)=\mathrm{span}\{\mathbf{1}_N\}$ and that $\mathrm{span}\{\mathrm{Ran}(L_a)\cup \mathrm{Ran}(L_b)\}\oplus \mathrm{span}\{\mathbf{1}_N \}=\mathbb{R}^N$.
	
\end{example}

\begin{figure}[h]
\centering\includegraphics[width=6cm]{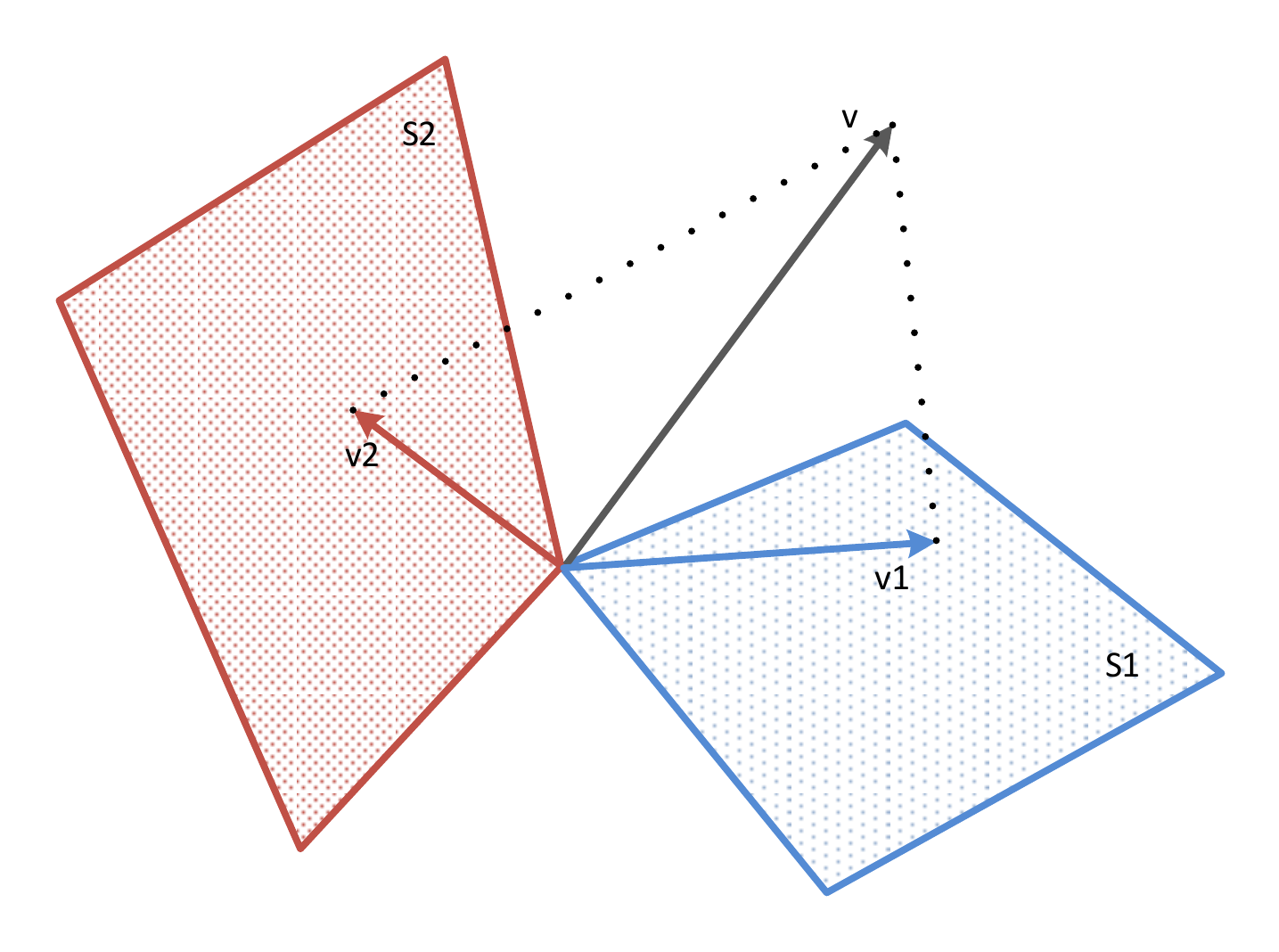}\caption{{\color{blue}Suppose we have two subspaces $s_1$ (the blue one) and $s_2$ (the red one), whose direct sum is the complementary space of synchronization manifold. Suppose from $[t_0,t_1)$, the norm of the projected state onto $s_1$ decreases while that of the projected state onto $s_2$ may increase. Then, from $[t_1,t_2)$, the norm of the projected state onto $s_2$ decreases while that of the projected state onto $s_1$ may increase. With this process continuing, it is possible to finally draw the conclusion that the projected state converges to zero if the decrease of the norm dominates its increase.}}\label{geometric-illustration-space-split}
\end{figure}

\begin{lemma}\label{convergence-in-subspace}
Given any linear time-invariant system
$\dot{x}=Hx,\;\;x(0)=x_0$
with $H\in\mathbb{R}^{n\times n}$,  the following two propositions hold.

$1$) Provided there exists a direct sum such that $\oplus_{j=1}^p S_j=\mathbb{R}^n$ with $S_i$ being of dimension $n_i$ and being invariant with respect to the linear mapping $H$. Construct the transformation matrix
$$
T=\bigg[\underbrace{v_1,\ldots,v_{n_1}}_{S_1},\underbrace{\ldots,\ldots}_{\ldots\ldots},\underbrace{v_{n-n_p+1},\ldots,v_n}_{S_p}\bigg],
$$
where $v_i$ is chosen as the basis vector of the invariant subspace $S_i,i=1,\ldots,p$.  Then, $\dot{\xi}_i=\tilde{H}_i\xi_i$ describes the evolution of $\mathrm{Prj}_{S_i}(x)$ \footnote{$\mathrm{Prj}_{S_i}(x)$ denotes the projection of $x$ into $S_i$. That $\dot{\xi}_i=\tilde{H}_i\xi_i$ describes the evolution of $\mathrm{Prj}_{S_i}(x)$ means $\|\xi_i(t)\|=\|\mathrm{Prj}_{S_i}(x(t)) \|$.} with $\xi_i$ defined by $\xi=[\xi_1^{\mathrm{T}},\ldots,\xi_p^{\mathrm{T}}]^{\mathrm{T}}=T^{-1}x$ and $\tilde{H}_i$ being the sub-matrix of
$
\tilde{H}=T^{-1}HT=\mathrm{diag}\left\{\tilde{H}_1,\ldots,\tilde{H}_p \right\}.
$

$2$) For any decomposition $x=x_1+x_2+\cdots+x_p$, where $x_i\in S_i$, which is of dimension $n_i$, and $ \oplus_{j=1}^p S_j=\mathbb{R}^n$,

{\color{blue} $\mathrm{i}$) if $\|x_i(t)\|\leq \psi\exp\{\gamma_i t\}\|x_i(0)\|$ for each $i=1,\ldots,p$, then there exists a positive constant $\Theta_{\mathbb{R}^n} (\psi)$ such that
$$
\|x(t)\|\leq \Theta_{\mathbb{R}^n}(\psi) \exp\{\max_i\gamma_i t\}\|x(0)\|;
$$

$\mathrm{ii}$) if $\|x(t)\|\leq \psi \exp\{\gamma t\}\|x_0\|,$ then  for each $i=1,\ldots,p$ and some positive constant $\Phi_{S_i}(\psi)$ depending on $\psi$, the decomposition of $\mathbb{R}^n$, and the state, one has
$$
\|x_i(t)\|\leq \Phi_{S_i}(\psi)\exp\{\gamma t\}\|x^{UB}_i(0)\|,
$$
where $\|x^{UB}_i(0)\|>0$ is a upper bound of $\|x_i(0)\|$;}

{\color{blue}$\mathrm{iii}$) given a different decompositions $x=y_1+\ldots+y_q$,   where $y_i\in \bar{S}_i$, which is of dimension $\bar{n}_i$, and $ \oplus_{j=1}^q \bar{S}_j=\mathbb{R}^n$, if $\|x_i(t)\|\leq \psi\exp\{\gamma_i t\}\|x_i(0)\|$, then $$\|y_i(t)\|\leq \Phi_{\bar{S}_i}(\Theta_S(\psi))\exp\{\max_i\gamma_i t\}\|y_i(0)\|,$$
where $S$ denotes the space of a direct sum of a subset of $\{S_1,\ldots,S_p\}$ that contains $\bar{S}_i$.
The subscripts of $\Theta$ and $\Phi$ indicate the space with where the corresponding variable evolves.}
\end{lemma}
It is worth pointing out that the conclusion drawn in $2$) of Lemma \ref{convergence-in-subspace} can also be applied to an autonomous nonlinear  system. The following remark specifies how to obtain $\Phi_{S_i}(\psi)$.
\begin{remark}	\label{rmk-notation}
$\Phi_{S_i}(\psi)$ in Proposition $\mathrm{ii}$) of 2) can be obtained as follows: Let $T=[v_1,\ldots,v_n]$, then $\Phi_{S_i}(\psi)$ can be chosen in such a way that {\color{blue}$\Phi_{S_i}(\psi)=\max_i\sigma_{\max}(M_iT^{-1})/\sigma_{\min}(T^{-1})\psi\cdot \max_i\|x_0\|/\|x^{UB}_i(0)\|$. Here, $M_i$ is a diagonal matrix having $n_i$ 1's from the ($\sum_{j=1}^{i-1}n_j+1$)-th to the ($\sum_{j=1}^{i}n_j$)-th entry. $\sigma_{\max}(\cdot)$ and $\sigma_{\min}(\cdot)$ denote the maximum and the minimum singular value of a matrix, respectively. It is worth pointing out that $\Phi_{S_i}(\cdot)$ relies on the decomposition of $\mathbb{R}^n$ and the initial choice of the upper bound $\|x^{UB}_i(0)\|$ to avoid the case $x_i(t_0)=0$, which is illustrated in the following example.} 
\end{remark}
{\color{blue}\begin{example}\label{illustrative-example}
See Fig.~\ref{geometric-illustration-space-split} for an example where $v_1$ and $v_2$ are assumed to be the projection states onto two eigenvectors of $H\in\mathbb{R}^{2\times 2}$. Let $u_1,u_2$ be the projection state onto another pair of basis vectors.  Suppose $u_1=\alpha v_1+\beta v_2$ with $\alpha,\beta\in\mathbb{R}$.  The increase of the norm of projection state onto $v_i$ is described by $\|v_i(t)\|\leq \psi_i \exp\{\lambda_i(t-t_1)\}\|v_i(t_1)\|$ with $\psi_i,\lambda_i>0$ for $i=1,2$. Then, the evolution of $u_1$ can be given by $\|u_1(t)\|\leq \max_i\psi_i \exp\{\max_i\lambda_i(t-t_1)\}\|u_1(t_1)\|$ if $\alpha$ and $\beta$ are nonnegative scalars.  Here, $\Phi_{S_i}(\psi)$ is bounded, uniformly with respect to the initial value of $u_1(t)$. However, if $\alpha>0,\beta<0$ and $\lambda_1>\lambda_2$, $\Phi_{S_i}(\psi)$ is not uniformly bounded. Moreover, it is possible that $\alpha v_1+\beta v_2=0$ in this case. This is why we choose an upper bound in ii) of proposition 2).
\end{example}}

\begin{lemma}\label{split-space}
Given subspaces $S_1,\ldots,S_p$, there exists a direct sum of $\bar{S}_1,\ldots,\bar{S}_{\bar{p}}$ such that $\oplus_{j=1}^{\bar{p}}\bar{S}_j=\mathrm{span}\{\cup_{j=1}^p S_j\}$ and $\bar{S}_j\subset S_k$ for some $k$.
\end{lemma}

\begin{example}[A Demonstration of Lemma \ref{split-space}]
	Consider the following two Laplacian matrices:
	\begin{align*}
	L_a=\begin{bmatrix}0&0&0&0\\-1&1&0&0\\0&0&0&0\\-1&0&0&1\end{bmatrix},\;\;L_b=\begin{bmatrix}0&0&0&0\\0&0&0&0\\-1&0&1&0\\0&0&-1&1\end{bmatrix}.
	\end{align*}
	$\mathrm{Ran}(L_a)$ is spanned by $v_1=[0,1,0,0]^{\mathrm{T}}$ and $v_2=[0,0,0,1]^{\mathrm{T}}$. While $\mathrm{Ran}(L_b)$ is spanned by $u_1=[0,0,1,0]^{\mathrm{T}}$ and $u_2=[0,0,0,1]^{\mathrm{T}}$. It is easy to verify that $\mathrm{Ran}(L_a)\cap \mathrm{Ran}(L_b)=\mathrm{span}\{[0,0,0,1]^{\mathrm{T}} \}$. One then obtains $S_1=\mathrm{span}\{[0,1,0,0]^{\mathrm{T}} \}$, $S_2=\mathrm{span}\{[0,0,0,1]^{\mathrm{T}} \}$, and $S_3=\mathrm{span}\{[0,0,1,0]^{\mathrm{T}}\}$.
	
\end{example}

{\color{blue}We shall explain in Sections  \ref{main-section} and \ref{sub-section-nonlinear} how to employ Lemmas \ref{semi-simple-zero-eigenvalue} to \ref{split-space} to conduct the analysis. Note that under joint connectivity condition, Lemmas \ref{kernel-intersect} and \ref{split-space} tell us that the space complementary to the synchronization manifold can be written into the direct sum of subspaces (see an illustration in Fig.~\ref{geometric-illustration-space-split}). These subspaces can be the eigenspaces of the nonzero eigenvalues of Laplacian matrices in linear case. They can also be subspaces in which the projection of the nonlinear self-dynamics still retains the Lipschitz property. Although the underlying topology might be disconnected at any time, in the subspaces mentioned above, the convergence property can be guaranteed. Then, one can invoke Lemma \ref{split-space} to extend the above convergence property to the complementary space of the synchronization manifold. }

\begin{lemma}\label{observability-extension}
The matrix pair $(C,A)$ is observable if and only if $(C,A-\Pi C)$ is observable for any compatible matrix $\Pi$.
\end{lemma}

 Finally, we introduce a useful result on how to construct the eigenvectors of zero eigenvalues of a Laplacian matrix. The following notations and definitions are needed:
	Let $R(j)$ be the set containing $j$ and all other nodes $i$ such that there exists a directed path from $j$ to $i$. A set $R$ of nodes in a graph will be called a \textit{reach} if it is a maximal reachable set; in other words, $R$ is a \textit{reach} if $R=R(i)$ for some $i$ and there is no $j$ such that $R(i)\subset R(j)$ (properly). For each reach $R_i$ of a graph, define the \textit{exclusive part} of $R_i$ to be the set $H_i=R_i\setminus( \cup_{j\neq i}R_j)$. Likewise, define the \textit{common part} of $R_i$ to be $C_i=R_i\setminus H_i$.
\begin{lemma}[cf. \cite{Caughman}]\label{kernel-space}
	Suppose $M=D-DS$, where $D$ is a nonnegative $N\times N$ diagonal matrix and $S$ is stochastic. Suppose ${\color{blue}\mathcal{G}}$ has $k$ reaches, denoted by $R_1,\ldots,R_k$, where we denote the exclusive and common parts of each $R_i$ by $H_i$ and $C_i$, respectively. Then the nullspace of $M$ has a basis $v_1,v_2,\ldots,v_k$ in  $\mathbb{R}^N$ whose elements satisfy:
		1) $v_i(s)=0$ for $s\notin R_i$;
	    2) $v_i(s)=1$ for $s\in H_i$;
		3) $v_i(s)\in (0,1)$ for $s\in C_i$; and
		4) $\sum_{j}v_j=\mathbf{1}_N$.
\end{lemma}
\begin{remark}\label{rmk-frobenius}
	It is obvious that $M$ is actually a Laplacian matrix of some non-negatively weighted digraph ${\color{blue}\mathcal{G}}$. To better understand Lemma \ref{kernel-space}, we write $M^i$, the Laplacian matrix of the $i$-th reach $R_i$, into the Frobenius normal  form \cite{Qin2015TNNLS}:
	\begin{align*}
	M^{i}=\begin{bmatrix}M^i_{11}&0&\cdots&0\\M^i_{21}&M^i_{22}&\cdots&0\\\vdots&\cdots&\ddots&\vdots\\M^i_{n_i,1}&\cdots&\cdots&M^i_{n_i,n_i}\end{bmatrix}
	\end{align*}
	where $M^i_{ss}$ corresponds to the subgraph ${\color{blue}\mathcal{G}^{i,s}}$ of ${\color{blue}\mathcal{G}}$ that is strongly connected for $s=1,\ldots,n_i.$ Moreover, $\mathcal{V}({\color{blue}\mathcal{G}^{i,1}})$, the node set of ${\color{blue}\mathcal{G}^{i,1}}$, belongs to the exclusive part of $R_i$ for $i=1,\ldots,k$. Otherwise, any node in ${\color{blue}\mathcal{G}^{i,s}}$ can be reached by the node outside, which contradicts the fact that $R_i$ is a reach. We call the nodes in $\mathcal{V}({\color{blue}\mathcal{G}^{i,1}})$ non-reachable nodes.
\end{remark}
\section{Evolution Of Systems Over Intervals Without Connectivity Requirement \label{main-section}}
In this section, with the help of the established results in Section \ref{sec:Main Results}, we will illustrate how the linear or Lipschitz-type nonlinear system evolves no matter how the communication graph is structured. To this aim, we first find the subspaces which can be the eigenspaces of the nonzero eigenvalues of Laplacian matrices in linear case and can also be subspaces in which the projection of the nonlinear self-dynamics still retains the Lipschitz property. Then, the analysis is performed with respect to these subspaces

%
The following analysis is performed in regard to the evolution of synchronization error. We start this section by the derivation of error system dynamics. For brevity, hereafter, we assume that $\sigma(t)$ is a periodic signal, which implies that $t_k-t_{k-1}=t_{k+1}-t_k=T$ and $\sigma(t)=\sigma(t+T)$. However, all the results obtained in this paper can be easily extended to the case that $\sigma(t)$ is not periodic.

\subsection{Error System Dynamics and State Space Decomposition of Linear System}
 Now, consider the commonly used synchronization error
 \begin{align}
 e=(\bar{\Delta}\otimes I_n)x=[(x_1-x_2)^{\mathrm{T}},\ldots,(x_1-x_N)^{\mathrm{T}}]^{\mathrm{T}},
 \end{align}
 where $\bar{\Delta}=\begin{bmatrix}\mathbf{1}_{N-1}&-I_{N-1}
 \end{bmatrix}.
 $ It is obvious that if $e=\mathbf{0}$, then $x_i=x_j$ for $i,j=1,\ldots,N$.

Now consider the time interval $[t_k,t_{k+1})$, which contains $m_k$ subintervals $[t_{k_0},t_{k_1}),\ldots,[t_{k_{m_k-1}},t_{k_{m_k}})$. Moreover, the union of communication graph over $[t_k,t_{k+1})$ contains a directed spanning tree. For the subinterval $[t_{k_{j}},t_{k_{j+1}})$, $0\leq j\leq m_k-1$, the compact form of the error system dynamics is
 \begin{align}\label{error-ssytem-dynamics}
 \dot{e}=\left(I\otimes A -\phi\tilde{L}_{\sigma(t_{k_j})}\otimes BK \right)e,
 \end{align}
 where $\tilde{L}_{\sigma(t_{k_j})}$ is the submatrix of
 $\Delta L_{\sigma(t_{k_j})}\Delta=\begin{bmatrix}0&*\\0&\tilde{L}_{\sigma(t_{k_j})}\end{bmatrix}$
 with
 $\Delta=[[1,0_{N-1}]^\mathrm{T},\bar{\Delta}^\mathrm{T}]^\mathrm{T}$ satisfying $\Delta\Delta=I_N$ and
  $*$ being an unspecified row vector. To perform our analysis, the following fact is important.

 \begin{lemma}\label{lma-shrink-space}
 	If the corresponding nonnegatively weighted graph of the Laplacian matrix $\sum_{j=0}^{m_k-1}L_{\sigma(t_{k_j})}$ contains a directed spanning tree, then
 	$$\cap_{j=0}^{m_k-1} \mathrm{Eig}\left(\tilde{L}_{\sigma(t_{k_j})}\right)=\{{\color{blue}\mathbf{0}}\},$$
    where $\mathrm{Eig}(\cdot)$ denotes the space spanned by the generalized eigenvectors associated with eigenvalue 0.
 \end{lemma}

From Lemma \ref{lma-shrink-space}, it can be known that   $\mathrm{span}\{\cup_{j=0}^{m_k-1} \mathrm{Ran}(\tilde{L}_{\sigma(t_{k_j})})\}=\mathbb{R}^{n(N-1)}$.
By Lemma \ref{split-space}, construct $d$ subspaces $S_i,i=1,\ldots,d$ such that $$\oplus_{j=1}^d S_j=\mathrm{span}\left\{\cup_{j=0}^{m_k-1} \mathrm{Ran}(\tilde{L}_{\sigma(t_{k_j})})\right\}.$$ Note that each $S_i\subset \mathrm{Ran}(\tilde{L}_{\sigma(t_{k_j})}\otimes I_n)$ for some $j$.
Now, the synchronization error $e$ can be decomposed in such a way that $e=e_1+\cdots+e_d$, where $e_i\in S_i$ for $i=1,\ldots,d$.
It suffices to show that $e_i,i=1,\ldots,d$ vanish as times approaches infinity in the following analysis to prove synchronization.


\subsection{Error System Dynamics and State Space Decomposition of Nonlinear System\label{nonlinear-section}}
{\color{blue} We discard the synchronization error $e$ used in the linear case due to the difficulty to maintain the Lipschitz property of the projection of nonlinear system dynamics onto $\mathrm{Ker}(L_{\sigma(t)}\times I_n)$ or $\mathrm{Ran}(L_{\sigma(t)}\times I_n)$. Instead, in order to capture the property of the evolution of system state in a certain subspace of $\mathbb{R}^{nN}$, we
construct a series of suitable error variables that evolve in subspaces where the projection of the nonlinear self-dynamics still retains the Lipschitz property. These error variables turn out to serve as the synchronization error in the sense that if all of them equal zero, then synchronization is realized.  The convergence analysis can then be performed with respect to each error variable.}

We first write $L_{\sigma(t)}$ into the Frobenius normal form \eqref{Frobenius-Form}:
\begin{align}\label{Frobenius-Form}
L_{\sigma(t)} =\begin{bmatrix}L^a_{\sigma(t)}&0\\
L^c_{\sigma(t)}&L^d_{\sigma(t)}\end{bmatrix},
\end{align}
where $L^a_{\sigma(t)}=\mathrm{diag}\left\{L_{\sigma(t),11},\ldots,L_{\sigma(t),\chi_{\sigma(t)}\chi_{\sigma(t)}}\right\}$,
$$
L^c_{\sigma(t)}=
\begin{bmatrix}
L_{\sigma(t),\chi_{\sigma(t)}+1,1}&\cdots&L_{\sigma(t),\chi_{\sigma(t)}+1,\chi_{\sigma(t)}}\\
\vdots&\ddots&\vdots\\
L_{\sigma(t),q1}&\cdots&L_{\sigma(t),q,\chi_{\sigma(t)}}
\end{bmatrix},
$$
and
$$
L^d_{\sigma(t)}=
\begin{bmatrix}
L_{\sigma(t),\chi_{\sigma(t)}+1,\chi_{\sigma(t)}+1}&\cdots&L_{\sigma(t),\chi_{\sigma(t)}+1,q}\\
\vdots&\ddots&\vdots\\
L_{\sigma(t),q,\chi_{\sigma(t)}+1}&\cdots&L_{\sigma(t),qq}
\end{bmatrix}.
$$


Note that fix $\chi_{\sigma(t)}< i\leq q$, there exists at least one $1\leq j\leq i$ such that $L_{\sigma(t),ij}\neq 0$.
$L_{\sigma(t),ii}(t)$ in \eqref{Frobenius-Form} corresponds to a subgraph ${\color{blue}\mathcal{G}_{\sigma(t)}^i}$ of ${\color{blue}\mathcal{G}_{\sigma(t)}}$ that is strongly connected for $i=1,\ldots,q$ and $\mathcal{V}({\color{blue}\mathcal{G}_{\sigma(t)}^i})$, the node set in ${\color{blue}\mathcal{G}_{\sigma(t)}^i}$, belongs to the exclusive part of the $i$-th reach $R_{\sigma(t),i}$ for $i=1,\ldots,\chi_{\sigma(t)}$.

With the concept provided in Lemma \ref{kernel-space}, we consider the following error variable
\begin{align}\label{error-nonlinear}
\delta_{\sigma(t)}(t)=x(t)-\sum_{j=1}^{\chi_{\sigma(t)}}  \left(\gamma_{\sigma(t),j}\beta_{\sigma(t),j}^{\mathrm{T}}\otimes I_n\right)x(t).
\end{align}
In \eqref{error-nonlinear}, $\gamma_{\sigma(t),j}=[\gamma_{\sigma(t),j}^1,\ldots,\gamma_{\sigma(t),j}^N]^{\mathrm{T}}$ is the right eigenvector of $L_{\sigma(t)}$ associated with eigenvalue zero corresponding to the reach $R_{\sigma(t),j}$ (see Lemma \ref{kernel-space}).
Moreover, $\beta_{\sigma(t),j}$ is defined such that $\beta_{\sigma(t),j}=[0,\ldots,0,v^{\mathrm{T}}_{\sigma(t),j},0,\ldots,0]^{\mathrm{T}}$ where $v_{\sigma(t),j}=[v_{\sigma(t),j}^1,\ldots,v_{\sigma(t),j}^{n_j}]^{\mathrm{T}}$, $n_j$ denotes the number of nodes in $\mathcal{V}(G_{\sigma(t)}^j)$, and $v_{\sigma(t),j}$  satisfies $v_{\sigma(t),j}^{\mathrm{T}} L_{\sigma(t),jj}=0$ and $v_{\sigma(t),j}^{\mathrm{T}} \mathbf{1}_{n_j}=1$. Hence, $\beta_{\sigma(t),j}^{\mathrm{T}} L_{\sigma(t)}=0$.

{\color{blue}There are two important properties of $\delta_{\sigma(t)}$. First, for $j=1,\ldots,\chi_{\sigma(t)},$
$
\left(\beta_{\sigma(t),j}^{\mathrm{T}}\otimes I_n\right)\delta_{\sigma(t)}
=0.
$
Second, $\mathrm{span}\{\delta_{\sigma(t)} \}\oplus \mathrm{span}\{\gamma_{\sigma(t),j}\otimes I_n,j=1,\ldots,\chi_{\sigma(t)} \}=\mathbb{R}^{nN}$. This is true because $\mathrm{dim}(\mathrm{span}\{\gamma_{\sigma(t),j}\otimes I_n,j=1,\ldots,\chi_{\sigma(t)} \})=n\chi_{\sigma(t)}$ and $\mathrm{dim}(\mathrm{span}\{\delta_{\sigma(t)} \})=nN-n\chi_{\sigma(t)}$. The latter can be known from the fact that } $$\left[\sum_{j=1}^{\chi_{\sigma(t)}}\left(\gamma_{\sigma(t),j}\beta_{\sigma(t),j}^{\mathrm{T}}\right)\right]^2=\sum_{j=1}^{\chi_{\sigma(t)}}\left(\gamma_{\sigma(t),j}\beta_{\sigma(t),j}^{\mathrm{T}}\right).$$

  The following lemma shows that if $\delta_{\sigma(t)}=\mathbf{0}$ for any $t$ under joint connectivity condition, then $x_i=x_j$ for $i\neq j.$

\begin{lemma}\label{synchronization-error}
Consider a collection of graphs $\{{\color{blue}\mathcal{G}_i}\}_{i=1}^p$, each of order $N$, whose union contains a directed spanning tree.  Then, $\cap_{j=1}^p\mathrm{span}(\{x|\delta_{j}=\mathbf{0}\})=\mathrm{span}\{\mathbf{1}_N\otimes u,\;u\in\mathbb{R}^n \}.$
\end{lemma}

 The compact form of the error system dynamics is given by
\begin{align}\label{error-ssytem-dynamics-nonlinear}
\dot{\delta}_{\sigma(t)}=&\;\dot{x}-\sum_{j=1}^{\chi_{\sigma(t)}}  \left(\gamma_{{\sigma(t)},j}\beta_{{\sigma(t)},j}^{\mathrm{T}}\otimes I_n\right)\dot{x} \notag\\
=&\;\mathbf{F}(x)-\phi(L_{\sigma(t)}\otimes \Gamma)x-\sum_{j=1}^{\chi_{\sigma(t)}} \left(\gamma_{{\sigma(t)},j}\beta_{{\sigma(t)},j}^{\mathrm{T}}\otimes I_n\right) \notag \\
\times&  \left[ \mathbf{F}(x)-\phi(L_{\sigma(t)}\otimes \Gamma)x\right] \notag \\
=&\left[I-\sum_{j=1}^{\chi_{\sigma(t)}}  \left(\gamma_{{\sigma(t)},j}\beta_{{\sigma(t)},j}^{\mathrm{T}}\otimes I_n\right)\right]\mathbf{F}(x) -\phi(L_{\sigma(t)}\otimes \Gamma) \notag \\
\times&\left[x-\sum_{j=1}^{\chi_{\sigma(t)}}  \left(\gamma_{{\sigma(t)},j}\beta_{{\sigma(t)},j}^{\mathrm{T}}\otimes I_n\right)x\right] \notag\\
=& M_{\sigma(t)} \mathbf{F}(x)-\phi(L_{\sigma(t)}\otimes \Gamma)\delta_{\sigma(t)},
\end{align}
where  $\mathbf{F}(x)=[f^{\mathrm{T}}(x_1),\ldots,f^{\mathrm{T}}(x_N)]^{\mathrm{T}}$ and $M_{\sigma(t)}=(I-\sum_{j=1}^{\chi_{\sigma(t)}}  (\gamma_{{\sigma(t)},j}\beta_{{\sigma(t)},j}^{\mathrm{T}}\otimes I_n))$. The third equality is obtained by the observation that
$$
(L_{\sigma(t)}\otimes \Gamma)\left[\sum_{j=1}^{\chi_{\sigma(t)}}  \left(\gamma_{{\sigma(t)},j}\beta_{{\sigma(t)},j}^{\mathrm{T}}\otimes I_n\right)\right]=0
$$
and $\beta_{\sigma(t),j}^{\mathrm{T}} L_{\sigma(t)}=0$. It is easy to verify that $M_{\sigma(t)}x(t)=\delta_{\sigma(t)}$.

\begin{example}
We will show the structure of $M_{\sigma(t)}\mathbf{F}(x)$ using a simple yet illustrative example. Consider a given communication graph in Fig.~\ref{Exp-Nonlinear}. There are two reaches $R_1=\{1,3 \}$ and $R_2=\{2,3\}$. Moreover, $H_i=\{3\},i=1,2$. Hence, $\gamma_1=[1,0,0.5]^{\mathrm{T}}$, $\beta_1=[1,0,0]^{\mathrm{T}}$, $\gamma_2=[0,1,0.5]^{\mathrm{T}}$, and $\beta_2=[0,1,0]^{\mathrm{T}}$. One can then obtain that
$$
M_{\sigma(t)}\mathbf{F}(x)=\begin{bmatrix}f(x_1)-f(x_1)\\f(x_2)-f(x_2)\\0.5(f(x_3)-f(x_1))+0.5(f(x_3)-f(x_2))\end{bmatrix}.
$$
\begin{figure}[htb]
\centering\includegraphics[width=6cm]{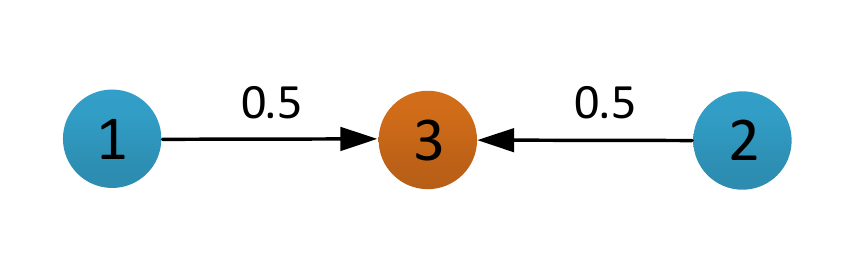}\caption{Directed graph with two reaches that share a common node.}\label{Exp-Nonlinear}
\end{figure}
\end{example}

By Lemma \ref{split-space} and Lemma \ref{synchronization-error},  find $d$ subspaces $S_i$ of $\mathrm{span}\{\delta_{\sigma(t)}\}$ for $t\in[t_{k},t_{k+1})$ such that $$\oplus_{j=1}^d S_i\oplus \mathrm{span}\{\mathbf{1}_N\otimes u,\;u\in\mathbb{R}^n \}=\mathbb{R}^{nN}.$$ Hence, one has $x=x_1+\ldots+x_{d+1}$ with $x_i\in S_i$ for $i=1,\ldots,{d}$ and $x_{d+1}\in\mathrm{span}\{\mathbf{1}_N\otimes u,\;u\in\mathbb{R}^n\}$. With respect to this decomposition, in the following proof, to guarantee synchronization, it suffices to show that {\color{blue}$\|x_1(t_{k+1})+\cdots+x_d(t_{k+1})\|< r \|x_1(t_{k})+\cdots+x_d(t_{k})\|$ with $0<r<1$ independent of $k$.}


\subsection{Evolution Property of the Projection of System State}
The following analysis is performed {\color{blue}with respect to} $[t_{k_j},t_{k_{j+1}})$ without loss of generality. Recall that $[t_k,t_{k+1})$ is supposed to contain $m_k$ sub-intervals, namely, $[t_{k_0},t_{k_1}),\ldots,[t_{k_{m_k-1}},t_{m_k})$ with $t_{k_0}=t_k$ and $t_{m_k}=t_{k+1}$, over each of which the communication graph remains unchanged.
\subsubsection{Linear System}
In this part, we analyze how $e(t)$ evolves within the fixed interval $[t_{k_j},t_{k_{j+1}})$. By Lemma \ref{convergence-in-subspace}, construct the transformation matrix $T_{\sigma(t)}$ whose column vectors are the generalized eigenvectors of $\tilde{L}_{\sigma(t)}$, and define $\xi=(T_{\sigma(t)}^{-1}\otimes I_n)e$. It then follows that
\begin{align}
\dot{\xi}_1=\left(I\otimes A-\phi\hat{L}_{\sigma(t)}\otimes BK \right)\xi_1, \label{xi-dynamics-1}
\end{align}
where
$\xi_1$-dynamics describes the evolution of $e$ restricted to the eigenspace of nonzero eigenvalues of $\tilde{L}_{\sigma(t)}$ and $ \hat{L}_{\sigma(t)}$ is a Hurwitz matrix.
One can then obtain from \eqref{xi-dynamics-1} that for $t\in[t_{k_{j}},t_{k_{j+1}})$
\begin{align}\label{evolution-range}
\xi_1(t)=\exp\left\{H^1_{{\sigma(t)}}\left(t-t_{k_j}\right) \right\}\xi_1(t_{k_j}),
\end{align}
with the matrix $H^1_{{\sigma(t)}}$ given by
$
H^1_{{\sigma(t)}}=I\otimes A-\phi\hat{L}_{{\sigma(t)}}\otimes BK.
$
Similarly, the evolution of $\xi_2$, which represents that of $e$ restricted to the kernel of $\tilde{L}_{\sigma(t)}$ during $t\in[t_{k_{j}},t_{k_{j+1}})$, is
\begin{align}
\dot{\xi}_2=(I\otimes A)\xi_2,
\end{align}
which in turn yields
\begin{align}\label{evolution-kernel}
\xi_2(t)=\exp\left\{ H^2_{\sigma(t)}(t-t_{k_j})\right\}\xi_2(t_{k_j}),
\end{align}
where $H^2_{\sigma(t)}=I\otimes A$.
\medskip
\subsubsection{Lipschitz-Type Nonlinear System}
In this part, we analyze the evolution of $x(t)$ is the subspaces constructed in Section \ref{nonlinear-section}.  When the evolution of $x$ is restricted to $\mathrm{span}\{\delta_{\sigma(t)} \}$, consider the Lyapunov function candidate $$V=\delta_{\sigma(t)}^{\mathrm{T}}\left(\Xi_{\sigma(t)}\otimes I_n\right)\delta_{\sigma(t)}$$ for \eqref{error-ssytem-dynamics-nonlinear},  where  the diagonal positive definite matrix $\Xi_{\sigma(t)}$ is defined such that for a positive constant $\alpha$
\begin{align}\label{choice-of-XI}
\delta_{\sigma(t)}^{\mathrm{T}}\left[\left(\Xi_{\sigma(t)} {L}_{\sigma(t)}+{L}_{\sigma(t)}^{\mathrm{T}}\Xi_{\sigma(t)}\right)\otimes I_n\right]&\delta_{\sigma(t)} \notag \\
>\alpha \delta_{\sigma(t)}^{\mathrm{T}}&\left[\Xi_{\sigma(t)}\otimes \Gamma\right]\delta_{\sigma(t)}.
\end{align}
The inequality \eqref{choice-of-XI} holds according to [Lemma 1, \cite{Qin2015TNNLS}] \footnote{The existence of positive constant $\alpha$ is guaranteed by the fact $\left(\beta_{\sigma(t),j}^{\mathrm{T}}\otimes I_n\right)\delta_{\sigma(t)}=0$. The details of how to calculate $\alpha$ are referred to \cite{Qin2015TNNLS} and omitted here for brevity.} and $(\beta_{\sigma(t),j}^{\mathrm{T}}\otimes I_n)\delta_{\sigma(t)}=0.$ $\alpha$ is well defined because $\sigma(t)$ takes finite values.
The derivative of $V$ along \eqref{error-ssytem-dynamics-nonlinear} then gives
\begin{align*}
\dot{V}=&2\delta_{\sigma(t)}^{\mathrm{T}}(\Xi_{\sigma(t)}\otimes I_n)\left(M_{\sigma(t)} \mathbf{F}(x)-\phi(L_{\sigma(t)}\otimes \Gamma)\delta_{\sigma(t)}\right) \\
= & 2\delta_{\sigma(t)}^{\mathrm{T}}(\Xi_{\sigma(t)}\otimes I_n)M_{\sigma(t)} \mathbf{F}(x) \\
-&\phi\delta_{\sigma(t)}^{\mathrm{T}}\left[\left(\Xi_{\sigma(t)}L_{\sigma(t)}+L^{\mathrm{T}}_{\sigma(t)}\Xi_{\sigma(t)}\right)\otimes \Gamma \right]\delta_{\sigma(t)}.
\end{align*}	
Consider the first term $\delta_{\sigma(t)}^{\mathrm{T}}(\Xi_{\sigma(t)}\otimes I_n)M_{\sigma(t)} \mathbf{F}(x)$. Since the inequality $\|M_{\sigma(t)}\mathbf{F}(x)\|\leq \bar{\rho}\|M_{\sigma(t)}x\|$ holds and $M_{\sigma(t)}x(t)=\delta_{\sigma(t)}$, one has $\|M_{\sigma(t)}\mathbf{F}(x)\|\leq \bar{\rho} \|\delta_{\sigma(t)}\|$.

With the above discussion, the derivative of $V$ along \eqref{error-ssytem-dynamics-nonlinear} turns into
\begin{align*}
\dot{V}\leq &
-\phi\alpha\delta_{\sigma(t)}^{\mathrm{T}}\left[\Xi_{\sigma(t)}\otimes \Gamma \right]\delta_{\sigma(t)}+2\bar{\rho}\lambda_{\max}(\Xi_{\sigma(t)})\left\|\delta_{\sigma(t)}\right\|^2  \notag\\
\leq & \left(-\phi\alpha\lambda_{\min}(\Gamma)+2\bar{\rho} c\right)V,
\end{align*}
where $c=\max_{t}\frac{\lambda_{\max}(\Xi_{\sigma(t)})}{\lambda_{\min}(\Xi_{\sigma(t)})}$. $c$ is well defined since $\sigma(t)$ has finite values. Through simple manipulations, one has
\begin{align}\label{Nonlinear-eqn-1}
&\|\delta_{\sigma(t)}(t)\| \notag \\
\leq &c^{1/2}  \exp\left\{ \left(-\frac{\phi\alpha}{2}\lambda_{\min}(\Gamma)+\bar{\rho} c\right)\left(t-t_{k_j}\right)\right\}
\left\|\delta_{\sigma(t)}(t_{k_j})\right\|,
\end{align}
for $t\in [t_{k_j},t_{k_{j+1}})$.

{\color{blue}On the other hand, one has the following
\begin{align}\label{error-perp}
\dot{x}(t)=\mathbf{F}(x(t))-\phi (L_{\sigma(t)}\otimes \Gamma)x(t).
\end{align}

Consider again the Lyapunov function $V=x^{\mathrm{T}}(\Xi_{\sigma(t)}\otimes I_n)x$. The derivative of $V$ along \eqref{error-perp} yields
\begin{align*}
\dot{V}&=2x^{\mathrm{T}}(\Xi_{\sigma(t)}\otimes I_n)\left(\mathbf{F}(x)-\phi (L_{\sigma(t)}\otimes \Gamma)x \right)\\
&\leq 2 \left[\rho\lambda_{\max}(\Xi_{\sigma(t)})+\phi\lambda_{\max}(\Xi_{\sigma(t)}L_{\sigma(t)}\otimes \Gamma)\right]\left\|x(t)\right\|^2 \\
&\leq 2(\rho c+c^{'}) V,
\end{align*}
where $c^{'}$ is defined by $c^{'}=\max_{t}\frac{\lambda_{\max}(\Xi_{\sigma(t)}L_{\sigma(t)}\otimes \Gamma+L^{\mathrm{T}}_{\sigma(t)}\Xi_{\sigma(t)}\otimes \Gamma)}{\lambda_{\min}(\Xi_{\sigma(t)})}$.
Since $\sigma(t)$ takes finite values, $c^{'}$, $\lambda_{\max}(\Xi_{\sigma(t)})$, and $\lambda_{\max}(L_{\sigma(t)}\otimes \Gamma)$ are well defined.}
It is then obtained that
\begin{align}\label{Nonlinear-eqn-2}
\|x(t)\|
\leq c^{1/2}  \exp\left\{\left(\rho c+c^{'}\right)\left(t-t_{k_j}\right)\right\} \times \left\|x(t_{k_j})\right\|.
\end{align}
One can then determine the evolution of $x$ restricted to $ \mathrm{span}\{\gamma_{\sigma(t),j},j=1,\ldots,\chi_{\sigma(t)} \}$ according to Lemma \ref{convergence-in-subspace}, which will be shown in the next section.
\section{Synchronization under Joint Connectivity Condition\label{sub-section-nonlinear}}
We illustrate in this section how the synchronization problem formulated in Section \ref{subsec:problem formulation} can be solved for linear system \eqref{linear-system-dynamics} and Lipschitz-type nonlinear system \eqref{nonlinear-system-dynamics} communicating over switching topology which is jointly connected.

General results are established in this section first, which provide sufficient algebraic conditions. Then, the applications of the newly-established results to some special cases to gain more insights are discussed.
\subsection{Inter-Connected Linear Systems}
{\color{blue}Before introducing our first main result, we claim some notations. $\hbar^i_{\sigma(t)}(\cdot)=\Phi_{\sigma_1(t)}(\cdot)$ if
$S_i\subset\mathrm{Ran}(L_{\sigma}(t)\otimes I_n)$; otherwise $\hbar^i_{\sigma(t)}(\cdot)=\Phi_{\sigma_2(t)}(\Theta_{\sigma_3(t)}(\cdot))$, where
$\Phi_{\sigma_2(t)}$ and $\Theta_{\sigma_1(t)}$ are given in Lemma \ref{convergence-in-subspace} and Remark \ref{rmk-notation}. $\sigma_1(t)$ indicates
the decomposition of $\mathrm{Ran}(L_{\sigma}(t)\otimes I_n)$ which includes $S_i$ as a subspace. $\sigma_2(t)$ and $\sigma_3(t)$ represent the
decompositions of the complementary  space of synchronization manifold, which include $S_i$ and $\mathrm{Ker}(L_{\sigma}(t)\otimes I_n)$ as subspaces, respectively}. {\color{blue}Moreover, we write
$$
\left\|\exp\left\{H^{ID}_{{\sigma(t)}}\right\}\right\|\leq\upsilon^{ID}_{\sigma(t)}\exp\left\{\xi^{ID}_{\sigma(t)} \right\}
$$
where $ID=1,2$. If $ID=1$, then
$H^1_{{\sigma(t)}}=I\otimes A-\phi\hat{L}_{{\sigma(t)}}\otimes BK$ and
$\hat{L}_{\sigma(t)}$ is Hurwitz which is defined in the same way as $\tilde{H}_i$ in Lemma \ref{convergence-in-subspace} corresponding to $\mathrm{Ran}(\tilde{L}_{\sigma(t)})$; if $ID=2$, then
$H^2_{{\sigma(t)}}=I\otimes A.$ Furthermore, let $\psi^i_{\sigma(t)}=\upsilon^1_{\sigma(t)}$ and $\lambda^i_{\sigma(t)}=\xi^1_{\sigma(t)}$ if $S_i\in\mathrm{Ran}(\tilde{L}_{\sigma(t)}\otimes I_n)$; $\psi^i_{\sigma(t)}=\upsilon^2_{\sigma(t)}$ and $\lambda^i_{\sigma(t)}=\xi^2_{\sigma(t)}$ otherwise.}

\begin{theorem}\label{Tho-Linear}
Considering the linear inter-connected system \eqref{linear-system-dynamics}, under Assumptions 1 and \ref{linear-assumtion}, synchronization is reached {\color{blue}if there exists a sufficiently small $\gamma\in(0,1)$ such that
\begin{align}\label{Tho-Condition}
\sum_{j=0}^{m_k-1}\left[\ln\left(\hbar^i_{\sigma(t_{k_j})}\left(\psi^i_{\sigma(t_{k_j})}\right)\right)+\lambda^i_{\sigma(t_{k_j})}\left(t_{k_{j+1}}-t_{k_j}\right)\right] < \ln \gamma
\end{align}
holds for any $k=1,2,\ldots$ and any $1\leq i\leq d$}.
\end{theorem}
\begin{IEEEproof}
Bearing \eqref{evolution-range} and \eqref{evolution-kernel} in mind,  when $S_i\subset \mathrm{Ran}\{\tilde{L}_{\sigma(t)} \}$, one has
\begin{align}
\left\|e_i(t) \right\|
\leq\hbar^i_{\sigma(t)}\left( \psi^i_{\sigma(t)} \right)\exp\left\{\lambda^{i-}_{\sigma(t)} \left(t-t_{k_j}\right) \right\}
\times \|e_i^{UB}(t_{k_j})\|,
\end{align}
where  $\hbar^i_{\sigma(t)}( \psi^i_{\sigma(t)})$ is given right before Theorem \ref{Tho-Linear} and $\|e_i^{UB}(t_{k_j})\|$ is an upper bound of $\|e_i(t_{k_j})\|$ at $t_{k_j}$. Here $\|e_i^{UB}(t_k)\|$ can be chosen as
$$
\max_i\sigma_{\max}(M_iT^{-1})/\sigma_{\min}(T^{-1})\|e(t_k)\|
$$
according to Lemma \ref{convergence-in-subspace} and Remark \ref{rmk-notation} ($T$ and $M_i$ are also defined according to the lemma and the remark).     Moreover, we use  $\lambda^{i+}_{\sigma(t)}$ ($\lambda^{i-}_{\sigma(t)}$) to denote that $\lambda^i_{\sigma(t)}>0$ ($\lambda^i_{\sigma(t)}<0$) in order to indicate the sign of $\lambda^i_{\sigma(t)}$.
Note that $\lambda^{i-}_{\sigma(t)}<0$ can be guaranteed by choosing appropriate feedback matrix $K$ such that $H^1_{\sigma(t)}$ is Hurwitz. The choice of $K$ is feasible since $\hat{L}_{\sigma(t)}$  is  Hurwitz.

Using Lemma \ref{convergence-in-subspace} when $S_i\not\subset \mathrm{Ran}(\tilde{L}_{\sigma(t)})$, the following inequality holds
\begin{align}\label{bounded-increase}
\left\|e_i(t)\right\|&\leq \hbar^i_{\sigma(t)}\left(\psi^i_{\sigma(t)}\right)\exp\left\{\lambda^{i+}_{\sigma(t)} \left(t-t_{k_j}\right) \right\} \times \|e_i^{UB}(t_{k_j})\|.
\end{align}
Invoking again Lemma \ref{convergence-in-subspace}  and  \eqref{evolution-kernel}, $\lambda^{i+}_{\sigma(t)}=\lambda^{i+}$ (independent of switching) is determined by the matrix $I\otimes A$.

Then, by recursion, one arrives at that
\begin{align}
\|e_i(t_{k+1})\|
\leq& \exp\Bigg\{\sum_{j=0}^{m_k-1}\ln\left(\hbar^i_{\sigma(t)}\left(\psi^i_{\sigma(t_{k_j})}\right)\right)\notag\\ 
&+\lambda^i_{\sigma(t_{k_j})}\left(t_{k_{j+1}}-t_{k_j}\right) \Bigg\}\|e_i^{UB}(t_k)\|.
\end{align}
   Hence, there exists a positive constant $\bar{\gamma}<1$ such that
$$
\|e(t_{k+1}) \| \leq \bar{\gamma} \|e(t_k)\|
$$
if the following inequality holds
\begin{align}\label{key-condition}
\sum_{j=0}^{m_k-1}\left[\ln\left(\hbar^i_{\sigma(t)}\left(\psi^i_{\sigma(t_{k_j})}\right)\right)+\lambda^i_{\sigma(t_{k_j})}\left(t_{k_{j+1}}-t_{k_j}\right)\right] < \ln \gamma,
\end{align}
where $\gamma<\frac{1}{N\max_i\sigma_{\max}(M_iT^{-1})/\sigma_{\min}(T^{-1})}$. This completes the proof by the observation that any $e$ can be written as $e=e_1+\ldots+e_i+\ldots+e_d$ with $e_i\in S_i$.
\end{IEEEproof}

\begin{algorithm}[htb]
{\color{blue}  \caption{Justification of the Convergence Condition \eqref{Tho-Condition}}
  \label{alg:Framwork}
  \begin{algorithmic}[1]
    \Require The switching scheme of the communication graph, the system dynamics, and the initial state.

    \Ensure
      Whether condition \eqref{Tho-Condition} holds or not.
    \State Design $K$ such that $I\otimes A-\hat{L}_{\sigma(t)}\otimes BK$ is a Hurwitz matrix.
    \State Calculate $\upsilon^{ID}_{\sigma(t)}$, $\xi^{ID}_{\sigma(t)}$ for $ID=1,2$ according to
      $$
      \exp\left\{I\otimes A-\hat{L}_{\sigma(t)}\otimes BK \right\}\leq \upsilon^{1}_{\sigma(t)}\exp\left\{ \xi^{1}_{\sigma(t)}\right\}
      $$
      and
      $$
      \exp\Big\{I\otimes A \Big\}\leq \upsilon^{2}_{\sigma(t)}\exp\left\{ \xi^{2}_{\sigma(t)}\right\}.
      $$
    \State Determine $\psi^i_{\sigma(t)}$ in such a way that $\psi^i_{\sigma(t)}=\upsilon^1_{\sigma(t)}$ and $\lambda^i_{\sigma(t)}=\xi^1_{\sigma(t)}$ if $S_i\in\mathrm{Ran}(\tilde{L}_{\sigma(t)})$; $\psi^i_{\sigma(t)}=\upsilon^2_{\sigma(t)}$ and $\lambda^i_{\sigma(t)}=\xi^2_{\sigma(t)}$ otherwise.
    \State Calculate $\hbar^i_{\sigma(t)}(\psi^i_{\sigma(t)})$ for $i=1,\ldots,d$ according to the following criteria:
    $\hbar^i_{\sigma(t)}(\cdot)=\Phi_{\sigma_1(t)}(\cdot)$ if
$S_i\in\mathrm{Ran}(L_{\sigma}(t)\otimes I_n)$; otherwise $\hbar^i_{\sigma(t)}(\cdot)=\Phi_{\sigma_2(t)}(\Theta_{\sigma_3(t)}(\cdot))$, where
$\Phi_{\sigma_2(t)}$ and $\Theta_{\sigma_1(t)}$ are given in Lemma \ref{convergence-in-subspace} and Remark \ref{rmk-notation}. $\sigma_1(t)$ indicates
the decomposition of $\mathrm{Ran}(L_{\sigma}(t)\otimes I_n)$ which includes $S_i$ as a subspace. $\sigma_2(t)$ and $\sigma_3(t)$ represent the
decompositions of $\mathbb{R}^{n(N-1)}$ which include $S_i$ and $\mathrm{Ker}(L_{\sigma}(t)\otimes I_n)$, respectively.
    \State Verify the inequality $$\sum\nolimits_{j=0}^{m_k-1}\big[\ln\hbar^i_{\sigma(t_{k_j})}(\psi^i_{\sigma(t_{k_j})}))+\lambda^i_{\sigma(t_{k_j})}(t_{k_{j+1}}-t_{k_j})\big] < \ln \gamma$$
    where $\gamma<\frac{1}{N\max_i\sigma_{\max}(M_iT^{-1})/\sigma_{\min}(T^{-1})}$.
  \end{algorithmic}}
\end{algorithm}

{\color{blue}
In regard to the condition \eqref{Tho-Condition}, we have provided Algorithm \ref{alg:Framwork} such that one can justify whether it is satisfied. The crucial step is to determine the function $\hbar_{\sigma(t)}^i$, which requires the  information of system state according to the calculation of $\Phi_{\sigma_i(t)}(\cdot),i=1,2$ by Remark \ref{rmk-notation}. However, in some cases, we do not need the knowledge of system state and can execute Algorithm \ref{alg:Framwork} in an easy way:
\begin{itemize}
\item Collect all the basis vectors in eigenspaces of nonzero eigenvalues of the matrix $ L_{\sigma(t)}\otimes I_n,t\geq 0$. If any two basis vectors have a nonnegative inner product and the inner products of the system state $x(t)$ and all the basis vectors have the same sign, then $\hbar_{\sigma(t)}^i$ is bounded for any time (see Example \ref{illustrative-example} for a simple illustration). This is because the projected state onto any one of the basis vectors has a linear bounded growth or decrease. Moreover, all these projections have the same sign. It is then clear that $\hbar^i_{\sigma(t)}(\cdot)$ is uniformly bounded.
\end{itemize}

In the following case, the above constraints on basis vectors and the system state can be satisfied.
\begin{itemize}

\item  If the linear inter-connected system \eqref{linear-system-dynamics} is a positive system \cite{Farina2000}, then choose the initial state such that its entries are nonnegative. Since the linear inter-connected system \eqref{linear-system-dynamics} is a positive system, one can find the basis vectors, which have nonnegative entries, in the range space of $L_{\sigma(t)}\otimes I_n$. In this case, $\hbar^i_{\sigma(t)}(\psi^i_{\sigma(t)})=\max_t \psi^i_{\sigma(t)}$, which is clearly upper bounded (see Example \ref{illustrative-example} for a simple illustration). Please note that in the above case, $\|e_i^{UB}(t_k)\|$ is chosen to be $\|e_i(t_k)\|$.
\end{itemize}
Although the above constraints seem strict, as will be observed from Example \ref{sim-linear-case}, there is no simple criterion that can determine the collective behavior of the inter-connected linear system \eqref{linear-system-dynamics}. For example, even if the system matrix $A$ is marginally stable and $K$ is designed such that $A-BK$ is Hurwitz, synchronization cannot be attained no matter how strong the coupling strength is and how long the dwell time is.
}

\begin{example}\label{sim-linear-case}
	The communication network switches between the two graphs ${\color{blue}\mathcal{G}_a}$ and ${\color{blue}\mathcal{G}_{b}}$ shown in Fig.~\ref{Fig-Exp-2-2}. ${\color{blue}\mathcal{G}_a}$ (${\color{blue}\mathcal{G}_b}$) consists of four nodes and the union of ${\color{blue}\mathcal{G}_a}$ and ${\color{blue}\mathcal{G}_b}$ contains a directed spanning tree.
	The Laplacian matrices of ${\color{blue}\mathcal{G}_a}$ and ${\color{blue}\mathcal{G}_b}$ are respectively given as follows
\begin{align*}
L_a=\begin{bmatrix}
0&0&0&0\\-1.2&1.2&0&0\\
0&0&0&0\\
0&-0.7&0&0.7
\end{bmatrix},\;
L_b=\begin{bmatrix}
0&0&0&0\\0&0&0&0\\
-0.5&0&0.5&0\\
0&0&-1.3&1.3
\end{bmatrix}.
\end{align*}

\begin{figure}
\centering\includegraphics[width=8cm]{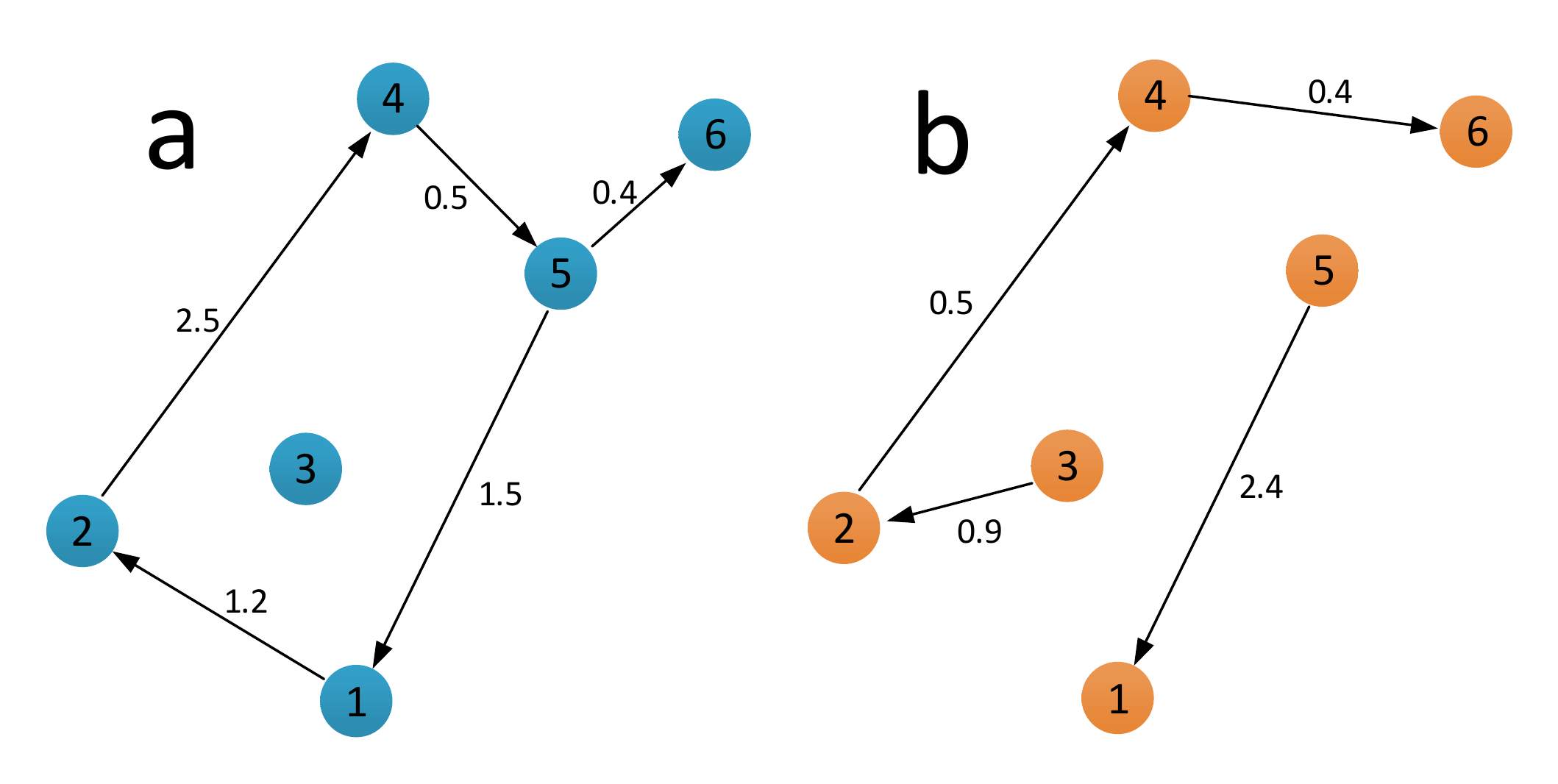}\caption{Two graphs (graphs $a$ and $b$) of a set of six nodes. The union of graphs $a$ and $b$ is a graph containing a directed spanning tree.}\label{Fig-Exp-2-1}
\end{figure}
\begin{figure}
\centering\includegraphics[width=8cm]{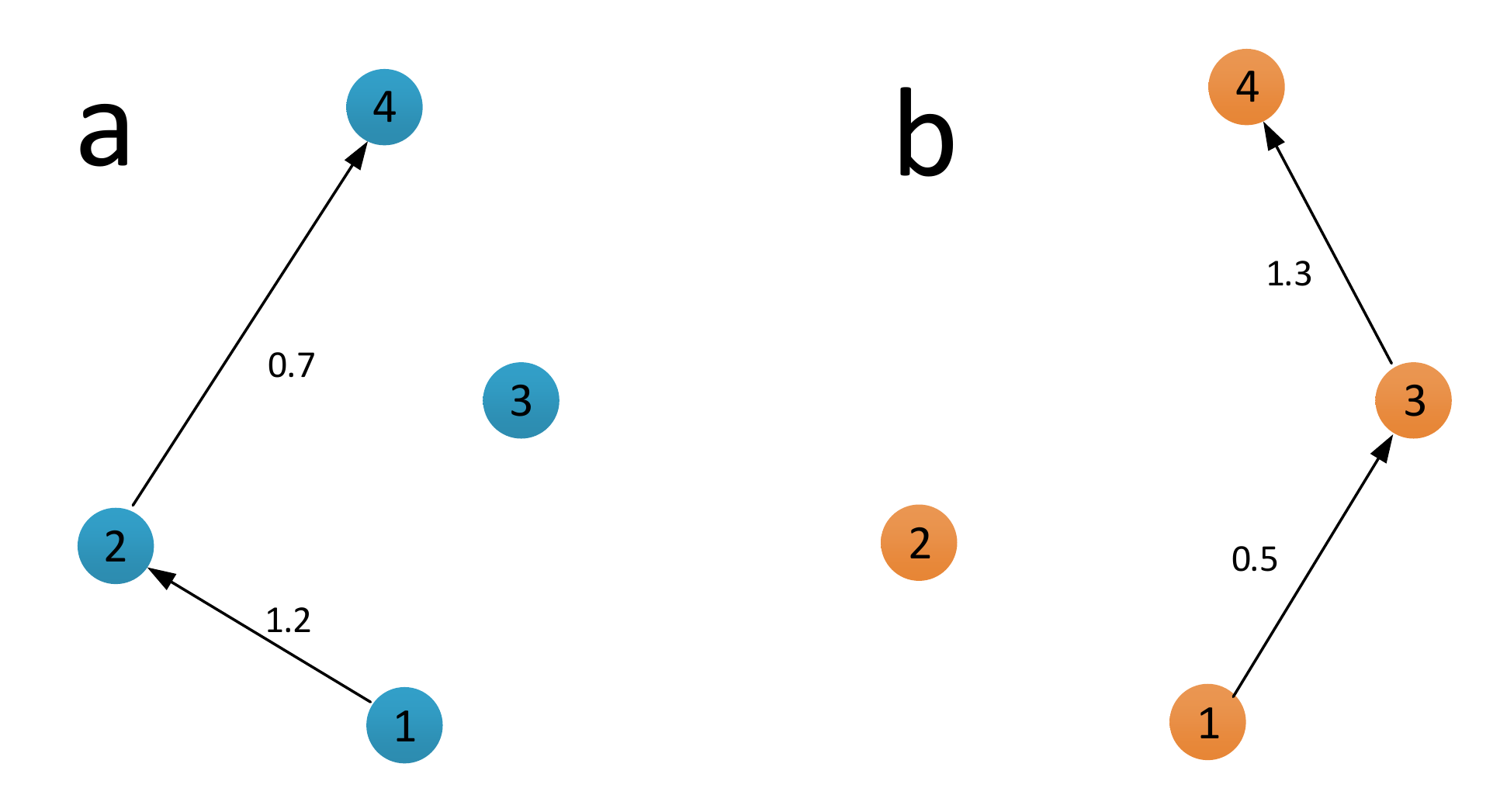}\caption{Two graphs (graphs $a$ and $b$) of a set of six nodes. The union of graphs $a$ and $b$ is a graph containing a directed spanning tree.}\label{Fig-Exp-2-2}
\end{figure}

For illustration, choose the following linear system and design the feedback matrix as
	\begin{align*}
	A={\begin{bmatrix}
		0& 1&0
		\\1 & 0 &0
		\\0&0&-2
		\end{bmatrix}},\; B={\begin{bmatrix}1&0\\0&1\\0&0
		\end{bmatrix}},\; K=\begin{bmatrix}1&0&0\\0&1&0\end{bmatrix}
	\end{align*}
	Obviously, $(A,B)$ is stabilizable.
$K$ is designed	to satisfy that $A-BK$ is Hurwitz and that the linear system is a positive one (because the off-diagonal entries of $I\otimes A-L_{\sigma(t)}\otimes BK$ is non-negative).
	Let the initial states be randomly chosen from $[0,50]\times[0,50]\times[0,50]\subset \mathbb{R}^{3}$. Moreover, let $T=1s$ be the dwell time of each communication graph.

It is known from $L_a$ and $L_b$ that the basis vectors in the eigenspaces of zero eigenvalue of them can be set as $v_1=[1,0,0,0]^{\mathrm{T}},v_2=[0,1,0,0]^\mathrm{T}$, and $v_3=[0,0,1,0]^{\mathrm{T}}$. By calculation, it is obtained that $\max_i\psi_i<1.4$. Moreover, $\lambda^i_{\sigma(t_0)}=0,i=1,2$, $\lambda^2_{\sigma(t_0)}=-3.29$, and $\lambda^2_{\sigma(t_1)}=-2.19$. According to Example \ref{illustrative-example}, let $\hbar^i_{\sigma(t)}(\psi^i_{\sigma(t)})(=\max_t\psi^i_{\sigma(t)})=1.4.$ It is easy to verify that condition \eqref{key-condition} is satisfied with $\gamma<1$.
It can be observed from Fig.~\ref{Linear-Fig-2-2-1} that synchronization is achieved asymptotically. Moreover, the increase of the coupling strength improve the convergence speed, which is shown in Fig.~\ref{Linear-Fig-2-2-2}.

 Suppose we consider the two graphs provided in Fig.~\ref{Fig-Exp-2-1} and choose the initial state from $[0,50]\times[0,50]\times[0,50]\subset \mathbb{R}^{3}$. Moreover set
$$
A=\begin{bmatrix}
		-1& 1&0
		\\1 & -1 &0
		\\0&0&-2
\end{bmatrix},\;
B=\begin{bmatrix}1&0\\0&0\\0&0
		\end{bmatrix},\;
 K=\begin{bmatrix}1&1&0\\0&1&0\end{bmatrix}.
$$
Although the union of the two graphs contains a directed spanning tree, $A$ is marginally stable, and $A-BK$ is Hurwitz, no matter how strong the coupling strength is or how long the dwell time is, synchronization cannot be achieved as shown in Fig.~\ref{Linear-Fig-2-1}. This is possibly because there are no basis vectors with nonnegative entries in the eigenspace of nonzero eigenvalues of the Laplacian matrices $L_a$ and $L_b$. Hence, $\hbar^i_{\sigma(t)}(\cdot)$ might not be bounded.
\begin{figure}
\centering\includegraphics[width=8cm]{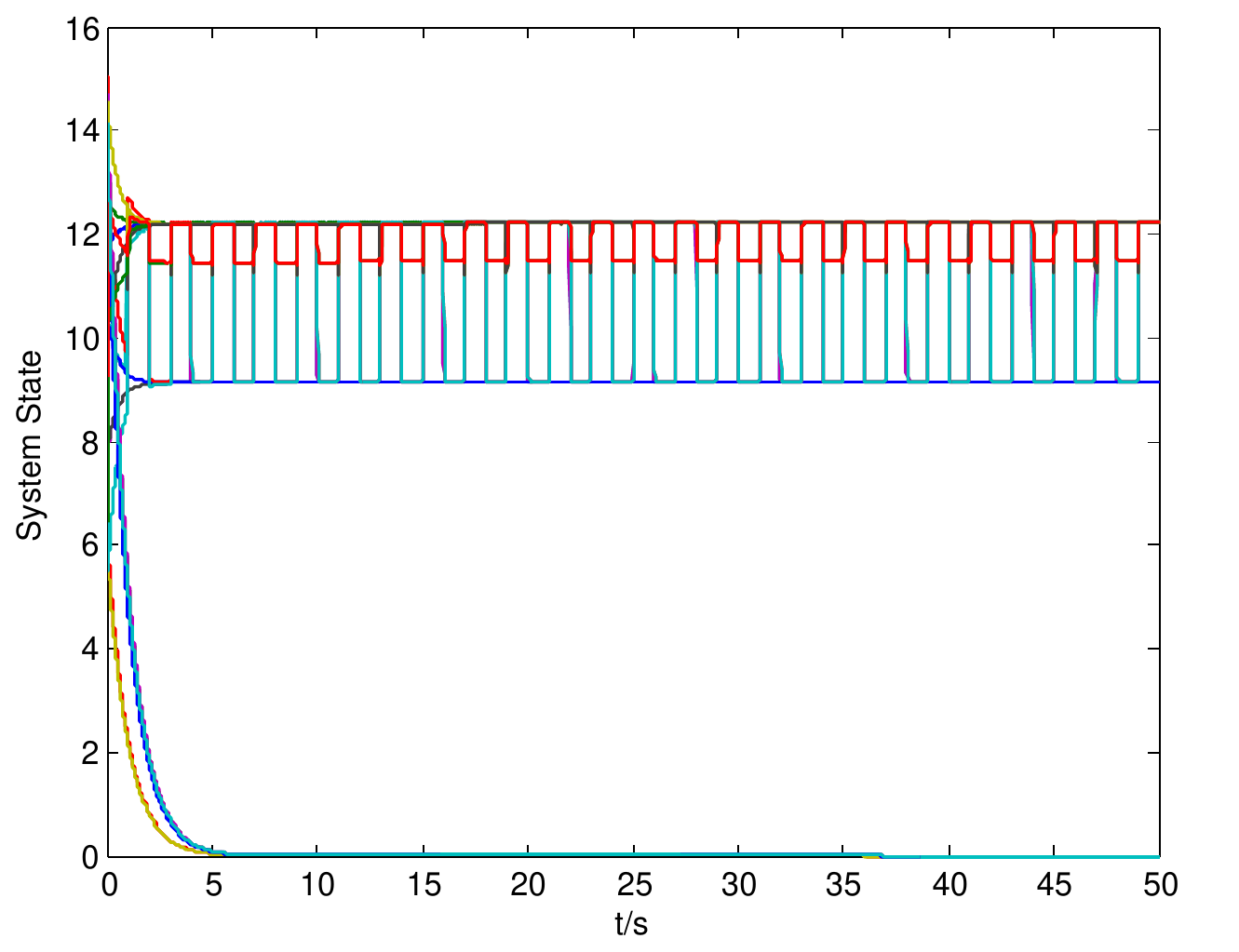}\caption{Evolution trajectories of system state with coupling strength $\phi=50$.}\label{Linear-Fig-2-1}
\end{figure}
\begin{figure}
\centering\includegraphics[width=8cm]{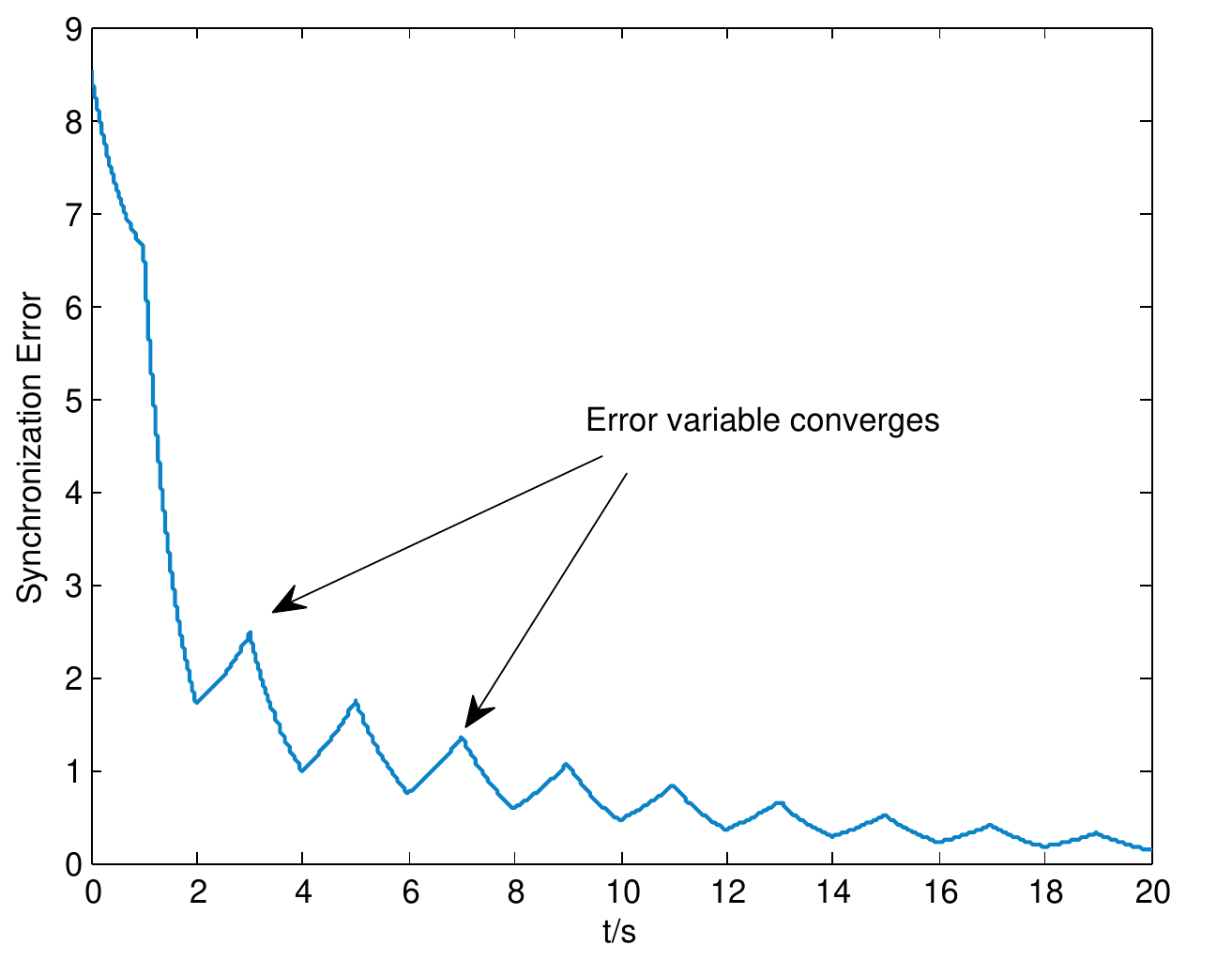}\caption{Evolution trajectories of synchronization error with coupling strength $\phi=5$.}\label{Linear-Fig-2-2-1}
\end{figure}
\begin{figure}
\centering\includegraphics[width=8cm]{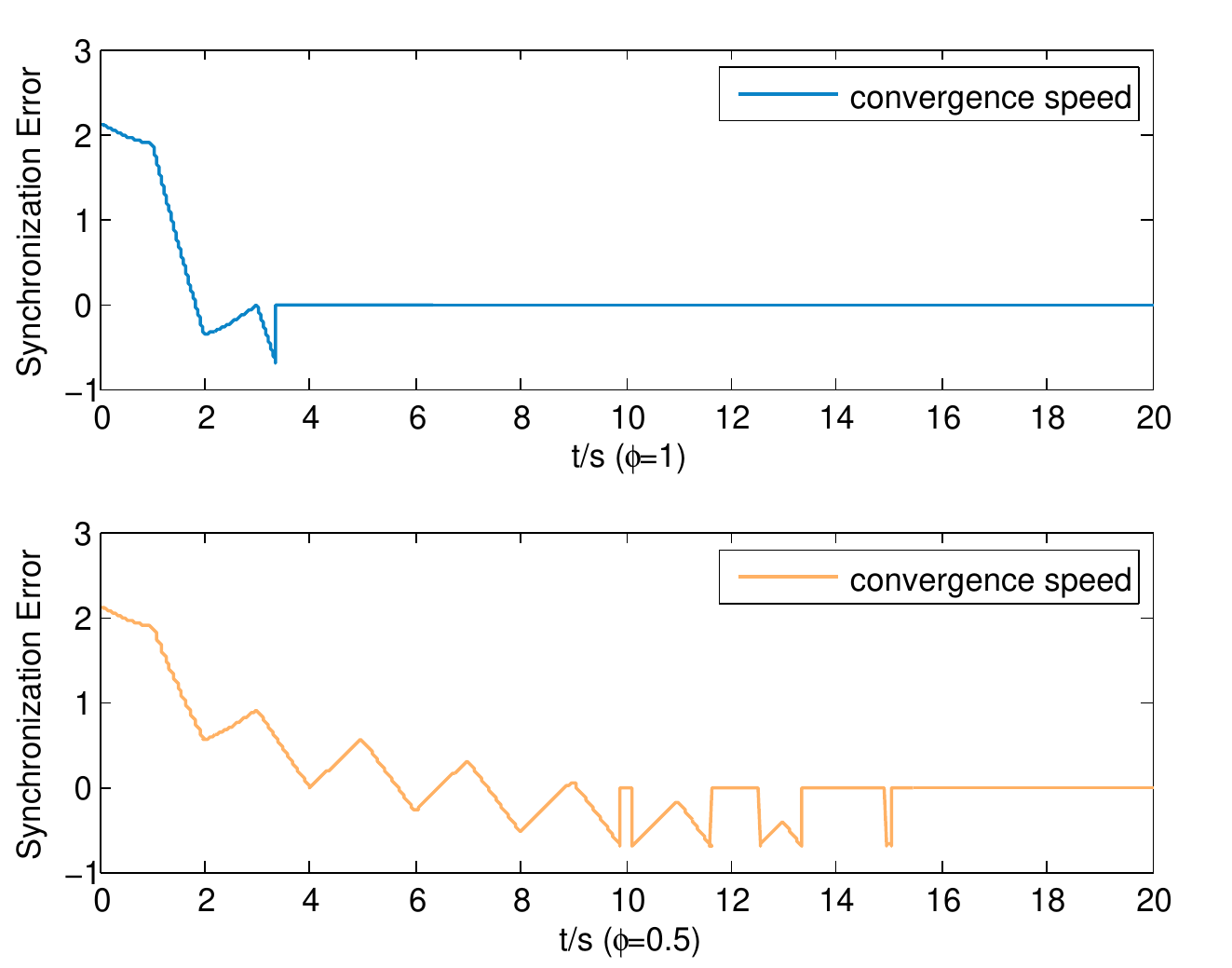}\caption{Convergence speed of synchronization error with $\phi=0.5$ and $\phi=1$, respectively.}\label{Linear-Fig-2-2-2}
\end{figure}
\end{example}

\smallskip
It is observed from \eqref{key-condition} that, the terms $\ln(\hbar^i_{\sigma(t)}(\psi^i_{\sigma(t)}))$, $\lambda^i_{\sigma(t)}$, and $(t_{k_{j+1}}-t_{k_j})$ are crucial for the convergence of \eqref{error-ssytem-dynamics}. Notice that $\lambda^i_{\sigma(t)}<0$ by choosing appropriate $K$ if the evolution of $e(t)$ is restricted to $\mathrm{Ran}(\tilde{L}_{\sigma(t)})$, while $\ln(\psi^i_{\sigma(t)})$ is sign indefinite. {\color{blue}There are two ways to address the issue that how to guarantee condition \eqref{key-condition}. The first one is to seek a smaller $\lambda^i_{\sigma(t)}<0$ to guarantee condition \eqref{key-condition}. The second one, on the other hand, is to permit a long dwell time, i.e., $T_{\min}$ is large enough. The following result illustrates the second case.}

{\color{blue}Before proceeding, according to Lemma \ref{convergence-in-subspace}, Remark \ref{rmk-notation}, and the above discussions, we make an additional assumption to facilitate the analysis.
\begin{assumption}\label{bounded-assumption}
$\Phi_{\sigma_1(t)}(\cdot)$ and $\Phi_{\sigma_2(t)}(\cdot)$ are uniformly bounded.
\end{assumption}}

\begin{theorem}\label{corollay-slow-varying}
Considering the linear inter-connected system \eqref{linear-system-dynamics},  suppose Assumptions 1, \ref{linear-assumtion}, and \ref{bounded-assumption} are satisfied and moreover $A$ is marginally stable \footnote{A matrix is said to be marginally stable if all its eigenvalues have non-positive real part and those with zero real part are semi-simple.}. Then, synchronization is reached if the dwell time is long enough, that is, $T_{\min}$ is sufficiently large.
\end{theorem}
Intuitively, to satisfy \eqref{key-condition} and hence guarantee the synchronization of linear system \eqref{linear-system-dynamics}, a large $T_{\min}$ is sufficient if $\lambda^i_{\sigma(t)}\leq 0$, which  holds because $A$ is marginally stable and $\hbar^i_{\sigma(t)}$ is uniformly bounded in Theorem \ref{corollay-slow-varying}.

Moreover, if $A$ is marginally stable and the communication graph is undirected, one can then conclude that the system state is non-increasing and bounded, so does the projection of the system state onto any eigenvector. Hence,  Assumption \ref{bounded-assumption} can be removed.
Recalling the system dynamics \eqref{linear-system-dynamics},
the following theorem illustrates that with mild restraint on system matrix $A$ and input matrix $B$, synchronization can be realized asymptotically for \eqref{linear-system-dynamics} over undirected networks.
Similar results have been established in  \cite{NiSCL2010} and \cite{SuTAC2012}. We provide a different yet simple and intuitive method based on our methodology. Moreover, our proof can easily be applied to arbitrary switching scheme (of undirected connected communication graph) discussed in \cite{ValcherAUTO2017}.

{\color{blue}We first record the closely related results on consensus over dynamic undirected topologies \cite{NiSCL2010,ValcherAUTO2017}.
\begin{lemma}[cf. \cite{NiSCL2010,ValcherAUTO2017}]\label{existing-result}
Consider the linear inter-connected system \eqref{linear-system-dynamics}. Suppose $(A,B)$ is stabilizable. 1) If $\mathcal{G}_{\sigma(t)}$ is jointly connected, and $K$ is designed as $K=B^{\mathrm{T}}P$ with $P$ given by
\begin{numcases}{}
A^{\mathrm{T}}P+PA\leq 0, \label{eqn-1}\\
A^{\mathrm{T}}P+PA-2\beta PBB^{\mathrm{T}}P+\beta I<0, \label{eqn-2}
\end{numcases}
where $\beta>0$ is determined by the communication graph, then leader-following consensus can be realized. 2) Assume $\mathcal{G}_{\sigma(t)}$ is connected all the time, and without loss of generality, $$A=\begin{bmatrix}A_u&\\&A_s\end{bmatrix}, \;B=\begin{bmatrix}B_u&\\&B_s\end{bmatrix},$$ where $A_s$ is Hurwitz and $A_u$ is a matrix with all its eigenvalues located in the closed right half-plane. Let $(s)=s^{d-1}+\alpha_{d-2}s^{d-2}+\cdots+\alpha_1s+\alpha_0$ be an arbitrary Hurwitz polynomial of degree $d-1$ and $K_u=k_d\bar{K}_u$ with $\bar{K}_u=[\alpha_0\;\cdots\;\alpha_{d-1}\;1]$. Then, there exists $k_d>0$ such that consensus can be realized asymptotically under arbitrary switching and $K=[K_u,0]$.
\end{lemma}

}
Now, we introduce the following technical assumption which is more relaxed than \eqref{eqn-1} and \eqref{eqn-2} in Lemma \ref{existing-result}.
\begin{assumption}[cf. \cite{SuTAC2012}]\label{asm-Lyapunov-negativeness}
	There exists a positive definite matrix $P$ such that
	\begin{align}\label{non-negative-condition}
	A^{\mathrm{T}}P+PA\leq 0,
	\end{align}
	and moreover $(B^{\mathrm{T}}P,A)$ is observable.
\end{assumption}
Intuitively, inequality \eqref{non-negative-condition} guarantees the non-increase of system state in the subspace associated with the kernel of the Laplacian matrix, while the observability of $(B^{\mathrm{T}}P,A)$ is imposed to ensure that the system state decreases strictly in the subspace associated with the range space of the Laplacian matrix.
\begin{theorem}\label{undirected-convergence}
Considering the linear inter-connected system \eqref{linear-system-dynamics} communicating over undirected switching graph,
under Assumptions \ref{connectivity-assumption}, \ref{linear-assumtion},  and \ref{asm-Lyapunov-negativeness},  synchronization  is realized.
\end{theorem}

\subsection{Inter-Connected Nonlinear Systems}
In this subsection, we extend the above results to the nonlinear case based on the lemmas  and the methodology esatablished in Section \ref{sec:Main Results}. First, we introduce our main result on Lipschitz-type nonlinear system.

\begin{theorem}\label{Tho-Nonlinear}
	Considering the nonlinear inter-connected system \eqref{nonlinear-system-dynamics}, under Assumptions \ref{connectivity-assumption} and \ref{bounded-assumption}, and suppose moreover for any $x\in\mathbb{R}^{nN}$ and some $\bar{\rho}>0$, it follows that {\color{blue}
	\begin{align}\label{extension-Lipschitz-condition}
	\left\|M_{\sigma(t)}\mathbf{F}(x)\right\|\leq \bar{\rho}\left\|M_{\sigma(t)}x\right\|,\;
	 \end{align}}
 then synchronization is reached in an exponential manner if the coupling strength $\phi$ is chosen to be sufficiently large.
\end{theorem}
\begin{IEEEproof}
Combining \eqref{Nonlinear-eqn-1} and \eqref{Nonlinear-eqn-2},  one arrives at
\begin{align}
{\color{blue}\|x_i(t_{k+1})\|}
\leq & \exp\Bigg\{ \frac{T_{\max}}{T_{\min}}\ln\left(\hbar^i_{\sigma(t)}\left(c^{1/2}\right)\right)
 -\frac{1}{2}\phi\int_{t=t_{k}}^{t_{k+1}}\alpha_{\sigma(t)}\mathrm{d}\tau \notag \\
  \times&\lambda_{\min}(\Gamma)+\max\left\{\rho c+c^{'},\bar{\rho}\right\}T_{\max}\Bigg\}
  {\color{blue}\left\|x_i^{UB}(t_{k})\right\|}, \notag
\end{align}
where $\hbar^i_{\sigma(t)}(\cdot)$ is defined in the same way as that in Theorem \ref{Tho-Linear}, $\|x_i^{UB}(t_{k})\|$ is the upper bound of $\|x_i(t_{k})\|$ at $t_k$ which can be given by
$$
\max_i\sigma_{\max}(M_iT^{-1})/\sigma_{\min}(T^{-1})\|x(t_k)\|
$$
according to Lemma \ref{convergence-in-subspace} and Remark \ref{rmk-notation} ($T$ and $M_i$ are also defined according to the lemma and the remark).
Moreover, $\alpha_{\sigma(t)}=\alpha>0$ if $x_i\in\mathrm{span}\{\delta_{\sigma(t)} \}$, $\alpha_{\sigma(t)}=0$ otherwise.
Hence, one can choose a sufficiently large $\phi$ such that
\begin{align*}
\frac{T_{\max}}{T_{\min}}\ln\left(\hbar_{\sigma(t)}^i\left(c^{1/2}\right)\right)
-&\frac{1}{2}\phi\int_{t=t_{k}}^{t_{k+1}}\alpha_{\sigma(t)}\mathrm{d}\tau \notag \\
\times \lambda_{\min}(\Gamma)T_{\min}+&\max\left\{\rho c+c^{'},\bar{\rho}\right\}T_{\max}<\ln \gamma,
 \notag
\end{align*}
for $\gamma<\frac{1}{N\max_i\sigma_{\max}(M_iT^{-1})/\sigma_{\min}(T^{-1})}$.
This completes the proof by the observation that any $x$ can be written as $x=x_1+\ldots+x_i+\ldots+x_d+x_{d+1}$ with $x_i\in S_i$ for $i=1,\ldots,d$.
\end{IEEEproof}
{\color{blue}
We have used Assumption \ref{bounded-assumption} to establish the theorem. Here is the detailed explanation of this assumption. Collect all the basis vectors of $\mathrm{span}\{\delta_{\sigma(t)}\},t\geq 0$. $\hbar^i_{\sigma(t)}(\cdot)$ is uniformly bounded if any two basis vectors have a nonnegative inner product and the initial value $x(0)$ is chosen such that the inner products of $\mathbf{F}(x)$ and all basis vectors have the same sign. This condition is true for an important class of positive systems \cite{Farina2000}.

Suppose the nonlinear inter-connected system \eqref{nonlinear-system-dynamics} is a positive system, which is true when the entries of $f(x_i)$ are nonnegative as long as those of $x_i$ are nonnegative. In this case, if the initial state is chosen such that all its components are nonnegative, then so does the system state at any time. Moreover, according to the properties of $\delta_{\sigma(t)}$, the basis vectors in $\mathrm{span}\{\delta_{\sigma(t)}\},t\geq 0$ might be chosen such that the components are nonnegative with those corresponding to nodes belonging to a nontrivial Reach (which has at least two nodes) being zero. That is, the basis vectors can take such a form $[0_{1\times (n_1+\ldots+n_{\chi_{\sigma(t)}})},*]^{\mathrm{T}}\in\mathbb{R}^{nN}$ with $*$ being a row vector of nonnegative entries. See the following example for an illustration. Please note that in the above case, $\|x_i^{UB}(t_k)\|$ is chosen to be $\|x_i(t_k)\|$.
\begin{example}
$f(x_i)$ can take a lot of forms such that the entries of $f(x_i)$ are nonnegative as long as those of $x_i$ are nonnegative, for example, all the coefficients of the terms in $f(x_i)$ are positive: $f(x_i)=2x_i+\exp\{-x_i\}$. Consider a simple Laplacian matrix and the associated matrix $M_{\sigma(t)}$ takes the following form
$$
L_a=\begin{bmatrix}0&0&0&0\\-1&1&0&0\\0&0&0&0\\0&0&0&0\end{bmatrix}\;\;
,
M_{\sigma(t)}=\begin{bmatrix}0&-1&0&0\\ -1&0&0&0\\0&0&0&0\\0&0&0&0\end{bmatrix}
$$
 The basis vectors can be chosen as $v_1=[1,0,0,0]^{\mathrm{T}}$ and $v_2=[0,1,0,0]^\mathrm{T}$. Then, $v_i^\mathrm{T} x\geq 0,i=1,2$ for any $x\in\mathbb{R}^4$ having nonnegative entries.
\end{example}
}
$M_{\sigma(t)}$ in \eqref{extension-Lipschitz-condition} is in fact a projection matrix by the observation that $M_{\sigma(t)}^2=M_{\sigma(t)}$.
It is noted that if $f(x)=Ax$,  condition \eqref{extension-Lipschitz-condition} holds. Moreover, if the communication graph is bidirected, condition \eqref{extension-Lipschitz-condition} can also be guaranteed, which will be shown in the next example and corollary.
%



	$\|M_{\sigma(t)}\mathbf{F}(x)\|\leq \bar{\rho}\|M_{\sigma(t)}x\|$
is imposed to deal with the influence the nodes of the common part of a reach receive from the nodes of exclusive parts. In the following corollary, we consider the special case that no common part exists in the communication graphs. It can be observed that  condition \eqref{extension-Lipschitz-condition} is {\color{blue}automatically satisfied}. {\color{blue}Note that if the communication graph is bidirected or undirected, then the common part of any reach is empty, and condition \eqref{extension-Lipschitz-condition} holds naturally.}
\begin{corollary}\label{nonlinear-corollary}
Considering the nonlinear inter-connected system \eqref{nonlinear-system-dynamics}, under Assumptions \ref{connectivity-assumption}, \ref{Lipschitz-assumtion}, and \ref{bounded-assumption}, and moreover if the common part of every reach is empty in each communication graph,
then synchronization is reached in an exponential manner if the coupling strength $\phi$ is chosen to be sufficiently large.
\end{corollary}

{\color{blue}\begin{example}
We show the structure of $M_{\sigma(t)}\mathbf{F}(x)$ for an undirected graph. Consider a given communication graph in Fig.~\ref{Exp-Nonlinear-5}. There is only one reach $R=\{1,2,3 \}$ in Fig.~\ref{Nonlinear-Fig-3}. Hence, $H=\{1,2,3\}$ and $\gamma=[1,1,1]^{\mathrm{T}}$, $\beta=[1/3,1/3,1/3]^{\mathrm{T}}$. One can then obtain that
\begin{align*}
\|M_{\sigma(t)}\mathbf{F}(x)\|=&\|M_{\sigma(t)}\left[\mathbf{F}(x)-\mathbf{F}((I-M_{\sigma(t)})x)\right]\|\\
\leq&\|M_{\sigma(t)}\|\left\|\begin{bmatrix}f(x_1)-f\left(\frac{1}{3}(x_1+x_2+x_3)\right)\\f(x_2)-f\left(\frac{1}{3}(x_1+x_2+x_3)\right)\\f(x_3)-f\left(\frac{1}{3}(x_1+x_2+x_3)\right)\end{bmatrix}\right\|\\
\leq& \rho\cdot \|M_{\sigma(t)}\| \cdot\|M_{\sigma(t)}x(t)\|.
\end{align*}
It is obvious that condition \eqref{extension-Lipschitz-condition} is satisfied.
\begin{figure}[htb]
\centering\includegraphics[width=6cm]{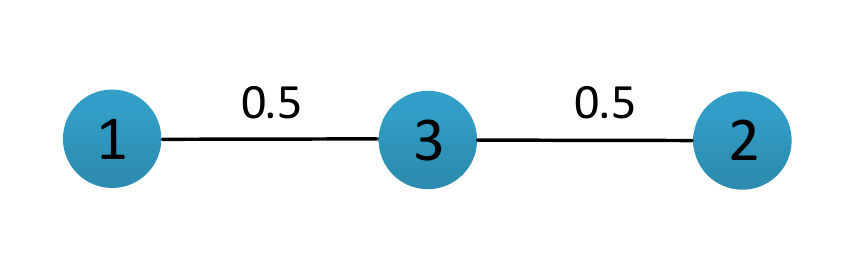}\caption{Undirected graph with two reaches that share a common node.}\label{Exp-Nonlinear-5}
\end{figure}
\end{example}}
{\color{blue}\begin{remark}
If $\mathcal{G}_{\sigma(t)}$ is bidirected \footnote{If in a graph, whenever $a_{ij}\neq 0$, there holds that $a_{ji}\neq 0$, then the graph is said to be bidirected. It is obvious that an undirected graph is bidirected.} all the time, then condition \eqref{extension-Lipschitz-condition} can be automatically satisfied.  This is because given any $t$, the reaches in $\mathcal{G}_{\sigma(t)}$ have no common part. Hence, the entry of the eigenvector $\gamma_{\sigma(t),j}$ is either 1 or 0.  One then knows that each block of $M_{\sigma(t)}\mathbf{F}(x)$ is zero or takes the form $f(x_i)-f(x_j)$. Then, the Lipschitz condition of $f(x)$ guarantees \eqref{extension-Lipschitz-condition}.
Ref. \cite{ChenAUTO2016} also deals with the synchronization analysis of Lipschitz-type nonlinear systems communicating over switching graph. It is concluded therein that if the coupling strength $\phi$ is sufficiently large, then the synchronization can be ensured. However, it is observed that $T_{\max}$ is also required to be sufficiently small such that $\phi T_{\max}\leq 1$. In contrast, our result removes the requirement on $T_{\max}$.
\end{remark}}
\begin{example}
	\begin{figure}
		\centering\includegraphics[width=8cm]{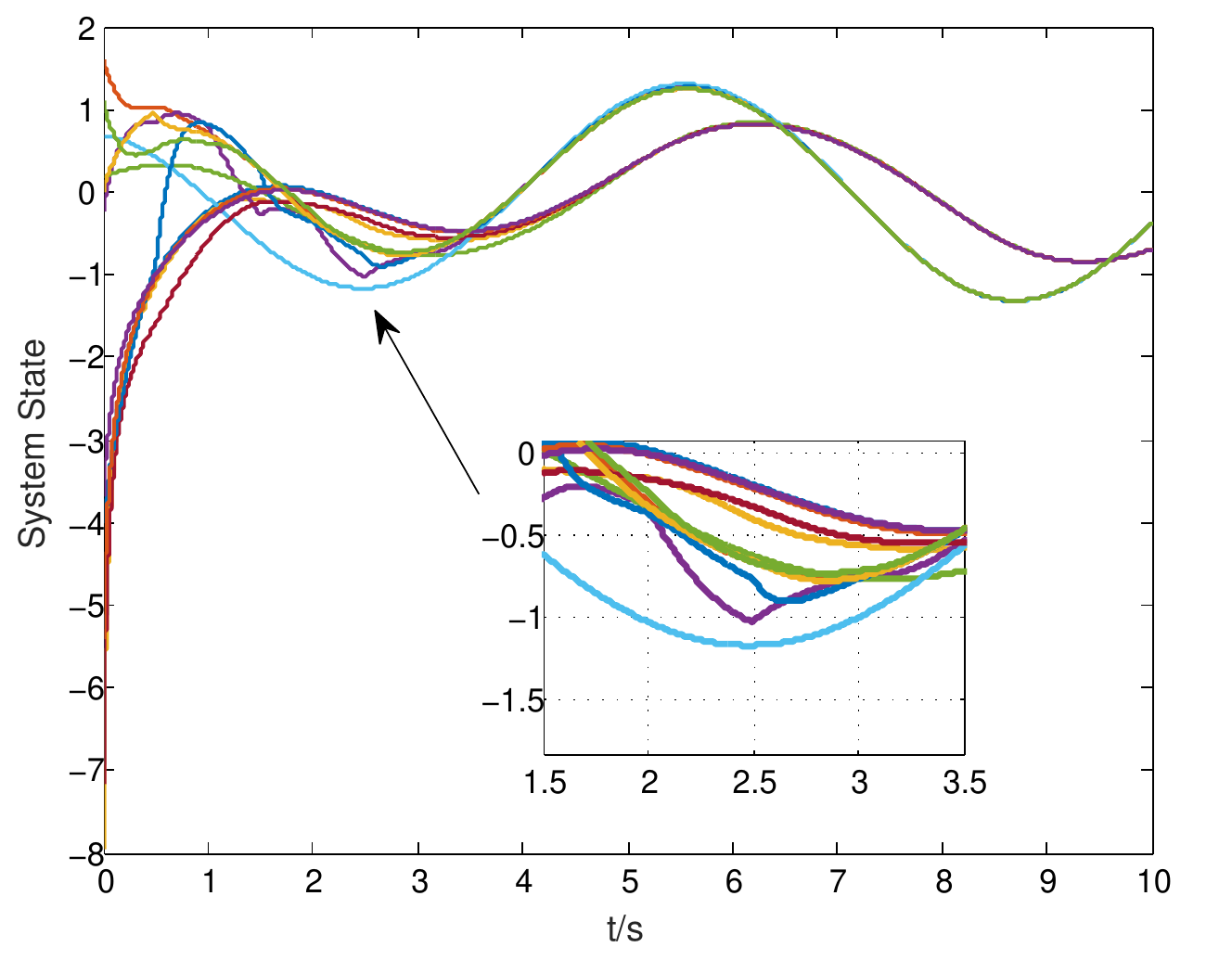}\caption{Evolution trajectories of the states of a collection of four nodes with $\phi=5$. It is observed from the trajectories that synchronization is asymptotically achieved.}\label{Nonlinear-Fig-3}
	\end{figure}
	Consider the interacting driving damped Van der Pol oscillators, for which the underlying coupling topology is shown in Fig.~\ref{Fig-Exp-2-2}, with $\Gamma=\mathrm{diag}\{0.5,0.9\}$ and $f(x)$ being given by
	\begin{equation}\notag
	f(x)=\left\{
	\begin{aligned}
	&x_2-1/3x_1^3-x_1,\\
	&-x_1+\sin(t).
	\end{aligned}\right.\,\,\,\,
	\end{equation}
	It is known that the Van der Pol oscillator is semi-contracting \cite{DeLellis2011}.
	The initial states is randomly chosen from $[-50,50]\times[-50,50]\times[-50,50]\subset \mathbb{R}^{3}$.
	Let $T=0.5s$ be the dwell time of each coupling topology. The switching of the coupling topology is therefore periodic. It can be observed from Fig. \ref{Nonlinear-Fig-3} that with $\phi=5$ synchronization is achieved asymptotically.
\end{example}

Next, a linearized technique will be exploited to reach a local result concerning the synchronization of inter-connected nonlinear system \eqref{nonlinear-system-dynamics}. Using of linearized method, as shall be observed later, can also relax the technical assumption on nonlinear function $f(\cdot)$.
 Rewrite the $i$-th element of $\mathrm{D}(M_{\sigma(t)}\mathbf{F}(x))$ as follows
\begin{align*}
\mathrm{D}(M_{\sigma(t)}\mathbf{F}&(x))_i=\mathrm{D}\left[f(x_i)-\sum_{j=1}^{\chi_{\sigma(t)}}\gamma^i_{\sigma(t),j}\left(\beta^{\mathrm{T}}_{\sigma(t),j}\otimes I_n\right) \mathbf{F}(x)\right]\\
=&\mathrm{D}\left[\sum_{j=1}^{\chi_{\sigma(t)}}\gamma^i_{\sigma(t),j}\left(\beta^{\mathrm{T}}_{\sigma(t),j}\otimes I_n\right) \left[\mathbf{1}_N\otimes f(x_i)-\mathbf{F}(x)\right]\right]\\
=&\mathrm{D}_f(x_i)\sum_{j=1}^{\chi_{\sigma(t)}}\gamma^i_{\sigma(t),j}\left(\beta^{\mathrm{T}}_{\sigma(t),j}\otimes I_n\right) \Big[\mathbf{1}_N\otimes x_i-x\Big],
\end{align*}
where $\mathrm{D}$ represents the  Jacobian  of $M_{\sigma(t)}\mathbf{F}\in\mathbb{R}^{nN}$. The second equality above holds by noting that $\sum_{j=1}^{\chi_{\sigma(t)}}\gamma_{\sigma(t),j}^k=1$ for $\forall k=1,\ldots,N$.
It thus follows that
\begin{align*}
\mathrm{D}\left(M_{\sigma(t)}\mathbf{F}\right)=\mathrm{D}_f\delta_{\sigma(t)},
\end{align*}
where $\mathrm{D}_f$ denotes the block diagonal matrix with each block being Jacobian matrix of the function $f$ at $x_i(t)$, i.e., $\mathrm{D}_f=\mathrm{diag}\{\mathrm{D}_f(x_1),\ldots,\mathrm{D}_f(x_N) \}$. By the definition of $\delta_{\sigma(t)}$, it follows that the linearized version of \eqref{error-ssytem-dynamics-nonlinear} is
\begin{align}\label{Nonlinear-eqn-3}
\dot{\delta}_{\sigma(t)}= \mathrm{D}_f \delta_{\sigma(t)}-\phi(L_{\sigma(t)}\otimes \Gamma)\delta_{\sigma(t)},
\end{align}
where we have used the approximation that $\mathbf{F}(x)=\mathrm{D}_fx$.
To obtain \eqref{Nonlinear-eqn-3}, it is required that the initial states are in the vicinity of a certain value.
Then, we have the following result concerning the stability of \eqref{Nonlinear-eqn-3}.
\begin{theorem}\label{nonlinear-local-result}
For the nonlinear inter-connected system \eqref{nonlinear-system-dynamics}, under Assumptions \ref{connectivity-assumption}, \ref{Lipschitz-assumtion}, and \ref{bounded-assumption}, and moreover if $\|\mathrm{D}_f\|<\rho$ for $x\in\mathbb{R}^{nN}$,
then synchronization is reached locally   if the coupling strength $\phi$ is chosen to be sufficiently large.	
\end{theorem}

It has been rigorously proven that there exists a coupling strength, say $\phi^*$, such that for any coupling strength $\phi>\phi^*$, synchronization for inter-connected Lipschitz-type nonlinear  systems can be ensured under certain conditions. Beyond this conclusion, one may further wonder whether one can find a $\phi_*$ such that  if $\phi<\phi_*$, synchronization cannot be ensured. In general, it is difficult to analyze the influence of the nonlinear function $f(\cdot)$ on synchronization, and hence the coupling strength $\phi_*$.  The following result illustrates that $\phi_*$ exists for nonlinear inter-connected systems with certain properties.

\begin{assumption}\label{asm-QUAD}
There exist two positive definite diagonal
matrices $Q=\mathrm{diag}\{q^{1},\ldots,q^{n}\}$ and
$\Sigma=\mathrm{diag}\{\delta^{1},\ldots,\delta^{n}\}$ such that for every
$x,y\in\mathbb{R}^{n}$, the below inequality holds:
\begin{equation}\label{QUAD-Inverse}
(x-y)^{\mathrm{T}}Q\big(f(x)-f(y)-\Sigma x+\Sigma y\big)\geq 0.
\end{equation}
\end{assumption}

Inequality \eqref{QUAD-Inverse} in Assumption \ref{asm-QUAD} is a converse version of the well-known \textbf{QUAD} condition \cite{DeLellis2011}. This condition encodes the anti-synchronization influence from the nonlinear function $f(\cdot)$. In addition, \eqref{QUAD-Inverse} holds if $\frac{\partial f}{\partial x}+\frac{\partial f}{\partial x}^{\mathrm{T}}$ is positive definite.

\begin{theorem}\label{Tho-Linear-critical-value}
Considering the nonlinear inter-connected system \eqref{nonlinear-system-dynamics}, under Assumptions \ref{connectivity-assumption} and  \ref{asm-QUAD}, synchronization cannot be reached if the coupling strength $\phi$ is chosen to be sufficiently small.
\end{theorem}


\section{Proof of The Main Results\label{Proof}}
\subsection{Technical Analysis for the Linear Case}

\begin{IEEEproof}[Proof of Theorem \ref{corollay-slow-varying}]
Since $A$ is marginally stable, it is easy to know that $\|\xi_2(t_{k_{j+1}})\|\leq \psi \|\xi_2(t_{k_{j}})\|$  for some constant $\psi$. Invoking Lemma \ref{convergence-in-subspace} and observing \eqref{evolution-range}, which has a negative growth rate $\lambda^i_{\sigma(t)}$ by choosing suitable $K$, there exists a $\hbar^i_{\sigma(t)}(\psi)>0$ such that
$$
\left\|e_i(t)\right\|\leq \hbar^i_{\sigma(t)}(\psi) \left\|e_i(t_{k_j}) \right\|,
$$
when $S_i\not\subset \mathrm{Ran}(\tilde{L}_{\sigma(t)})$.
To guarantee \eqref{key-condition}
it suffices to have
\begin{align}
\sum_{j=0}^{m_k-1}\ln\left(\hbar^i_{\sigma(t_{k_j})}\left(\psi^i_{\sigma(t_{k_j})}\right)\right)-\ln \gamma \leq -\lambda^i_{\sigma(t_{k_i})}T_{\min},
\end{align}
for sufficient small $0<\gamma<1$ if $S_i\subset \mathrm{Ran}(\tilde{L}_{\sigma(t)})$ when $t\in[t_{k_i},t_{k_{i+1}})$.
Hence, a lower bound of $T_{\min}$ is given by
$$
T_{\min}>\left(\sum_{j=0}^{m_k-1}\ln\left(\hbar^i_{\sigma(t_{k_j})}\left(\psi^i_{\sigma(t_{k_j})}\right)\right)-\ln \gamma\right)\bigg/-\lambda^i_{\sigma(t_{k_i})}.
$$
The proof is complete since $\hbar^i_{\sigma(t)}(\cdot)$ is uniformly bounded.
\end{IEEEproof}
\begin{IEEEproof}[Proof of Theorem \ref{undirected-convergence}]
It is noted that the communication graph for $t\in[t_{k_j},t_{k_{j+1}})$ is ${\color{blue}\mathcal{G}_{\sigma(t)}}$.
Now, let $v_{\sigma(t)}^1,\ldots,v_{\sigma(t)}^{N}$ be the orthogonal eigenvectors of $L_{\sigma(t)}$ with $v_{\sigma(t)}^N\in\mathrm{span}\{\mathbf{1}_N \}$.

In the following, $v_{\sigma(t)}^i$ are fixed during $[t_{k_j},t_{k_{j+1}})$. Then, $x$ can be further decomposed as
$
x=\alpha_{\sigma(t)}^1\bar{v}_{\sigma(t)}^1+\cdots+\alpha_{\sigma(t)}^{N-1}\bar{v}_{\sigma(t)}^{N-1}+\alpha_{\sigma(t)}^{N}\bar{v}_{\sigma(t)}^{N},
$	
where $\alpha_{\sigma(t)}^s\in\mathbb{R}$ and $\bar{v}_{\sigma(t)}^s=v_{\sigma(t)}^s\otimes \pi_{\sigma(t)}^s$ for $s=1,\ldots,N$.

Recall that the linear inter-connected system  \eqref{linear-system-dynamics} communicate over ${\color{blue}\mathcal{G}_{\sigma(t)}}$ during $[t_{k_j},t_{k_{j+1}})$. The evolution of $x_{\sigma(t)}^s(t)=\alpha_{\sigma(t)}^s\bar{v}_{\sigma(t)}^s(t)$ is then described by
\begin{align}\label{eqn-1}
\dot{x}_{\sigma(t)}^s=\left(I_N\otimes A\right)x_{\sigma(t)}^s, \;
\end{align}
for $v_{\sigma(t)}^s\in \mathrm{Ker}(L_{\sigma(t)})$, and
\begin{align}\label{eqn-2}
\dot{x}_{\sigma(t)}^s=\left[I_N\otimes (A-\phi\lambda_{\sigma(t)}^s BB^{\mathrm{T}}P)\right]x_{\sigma(t)}^s, \;
\end{align}
for $v_{\sigma(t)}^s\in \mathrm{Ran}(L_{\sigma(t)})$ by choosing $K=B^{\mathrm{T}}P$. $\lambda_{\sigma(t)}^s$ is the nonzero eigenvalue of $L_{\sigma(t)}$ associated with $v_{\sigma(t)}^s$.

Define $V_{\sigma(t)}^s=x_{\sigma(t)}^{s\mathrm{T}}(I_{N}\otimes P)x_{\sigma(t)}^s$. Taking the derivative of $V_{\sigma(t)}^s$ along \eqref{eqn-1} gives
\begin{align*}
\dot{V}_{\sigma(t)}^s=x_{\sigma(t)}^{s\mathrm{T}}\left[I_N\otimes \left(A^{\mathrm{T}}P+PA\right)\right]x_{\sigma(t)}^s\leq 0,
\end{align*}
which implies that $\|(I_N\otimes P^{\frac{1}{2}})x_{\sigma(t)}^s\|$ is non-increasing if $v_{\sigma(t)}^s\in \mathrm{Ker}(L_{\sigma(t)})$. Similarly, the derivative of $V_{\sigma(t)}^s$ along \eqref{eqn-2} yields
\begin{align*}
\dot{V}_{\sigma(t)}^s=x_{\sigma(t)}^{s\mathrm{T}}\left[I_N\otimes \left(A^{\mathrm{T}}P+PA-2\phi\lambda_{\sigma(t)}^s PBB^{\mathrm{T}}P\right)\right]x_{\sigma(t)}^s\leq 0.
\end{align*}

Consider the Lyapunov function candidate for \eqref{linear-system-dynamics} that is the sum of $V_{\sigma(t)}^s,s=1,\ldots,N$:
 $$V(t)=\sum\nolimits_{s=1}^{N} V_{\sigma(t)}^s=x^{\mathrm{T}}\left(I_N\otimes P \right)x,$$
 where $x=\sum_{s=1}^{N}x_{\sigma(t)}^s$.
The derivative of V along \eqref{linear-system-dynamics} yields
$
\dot{V}=\sum_{s=1}^{N} \dot{V}_{\sigma(t)}^s\leq 0.
$
Hence, $V(t)$ is non-increasing and converges to a nonnegative constant $V^*$ as time approaches infinity. Moreover, if $V_{\sigma(t)}^s=0$ during $[t_{k_j},t_{k_j+1}),\,\forall\, 0\leq j\leq m_k-1$ with $s$ satisfying $v_{\sigma(t)}^s\in\mathrm{Ran}(L_{\sigma(t)})$, then one obtains that  synchronization is achieved for \eqref{linear-system-dynamics} by Lemma \ref{kernel-intersect}. Now, we suppose that for some $1\leq s\leq N-1$ satisfying $v_{\sigma(t)}^s\in\mathrm{Ran}(L_{\sigma(t)})$ and $0\leq j\leq m_k-1$, $V_{\sigma(t)}^s(t_{k_j})\geq Z^*>0$ as $k\to \infty$ and show that contradiction will be incurred in the following.

Since $V_{\sigma(t)}^s(t_{k_j})\geq Z^*$,
$\|(I\otimes P^{1/2})x_{\sigma(t)}^s({t_{k_j}})\|\geq  \sqrt{Z^*}$. In addition, $\|(I\otimes P^{1/2})x_{\sigma(t)}^s(t_{k_j}) \|\geq \|(I\otimes P^{1/2})x_{\sigma(t)}^s(t_{k_{j+1}}) \|$ according to the negativity of the derivative of $V_{\sigma(t)}^s$.  Noting that $(B^{\mathrm{T}}P,A)$ is observable, by Lemma \ref{observability-extension}, it is known that $(B^{\mathrm{T}}P,A-\phi\lambda_{\sigma(t)}^s BB^{\mathrm{T}}P)$ is also observable.
One then has
\begin{align*}
&V_{\sigma(t)}^s(t_{k_{j+1}})-V_{\sigma(t)}^s(t_{k_j})\\
&\leq-2\phi\lambda_{\sigma(t)}^s x_{\sigma(t)}^{s\mathrm{T}}\left(t_{k_j}\right)\\
& \times\bigg[\int_{t_{k_j}}^{t_{k_{j+1}}}\exp\left\{I\otimes \left(A-\phi\lambda_{\sigma(t)}^s BB^{\mathrm{T}}P\right)(\tau-t_{k_j})\right\}^{\mathrm{T}}\left(I\otimes PB\right) \\
&~~ \times \left(I\otimes B^{\mathrm{T}}P\right) \exp\left\{I\otimes\left(A-\phi\lambda_{\sigma(t)}^s BB^{\mathrm{T}}P\right)(\tau-t_{k_j})\right\} \mathrm{d}\tau\bigg]\\
&~~~~\times x_{\sigma(t)}^s\left(t_{k_j}\right)\\
&\leq-2\phi\lambda_{\sigma(t)}^s x_{\sigma(t)}^{s\mathrm{T}}\left(t_{k_j}\right)Ox_{\sigma(t)}^s\left(t_{k_j}\right).
\end{align*}
The observability of $(B^{\mathrm{T}}P,A-\phi\lambda_{\sigma(t)}^s BB^{\mathrm{T}}P)$ implies that $O$ exists and is positive definite \cite{Antsaklis}. Moreover, since $\|x_{\sigma(t)}^s(t_{k_j})\|>\sqrt{Z^*}/\sigma_{\max}(P^{1/2})$ where $\sigma_{\max}(P^{1/2})$ denotes the maximum singular value of $P^{1/2}$, $V_{\sigma(t)}^s(t_{k_{j+1}})-V_{\sigma(t)}^s(t_{k_j})<-a^*$ for a sufficiently large $t_{k_j}$ and a positive constant $a^*$. This contradicts the fact that $V(t)$ is a Cauchy sequence. Hence, $V_{\sigma(t)}^s(t_{k_j})\to 0$ as $k\to \infty$, which in turn guarantees the achievement of synchronization for \eqref{linear-system-dynamics}.
\end{IEEEproof}

\subsection{Technical Analysis in the Nonlinear Case}

\begin{IEEEproof}[Proof of Theorem \ref{nonlinear-local-result}]
The proof follows that of Theorem \ref{Tho-Nonlinear}. We omit the detailed proof for brevity
%
\end{IEEEproof}
\begin{IEEEproof}[Proof of Theorem \ref{Tho-Linear-critical-value}]
The proof is with respect to synchronization error $e$ rather than $\delta_{\sigma(t)}$. It is easy to obtain that
\begin{align}\label{error-nonlinear-new}
\dot{e}=\mathbf{f}(x)-\phi\left(\tilde{L}_{\sigma(t)}\otimes \Gamma\right)e,
\end{align}
where $\mathbf{f}(x)=[(f(x_1)-f(x_2))^{\mathrm{T}},\ldots,(f(x_1)-f(x_N))^{\mathrm{T}}]^{\mathrm{T}}$.
Consider the Lyapunov function candidate
$
V(t)=e^{\mathrm{T}}(\Xi\otimes Q)e
$
for the error system dynamics \eqref{error-nonlinear-new}, where $\Xi$ is a positive diagonal matrix and $Q$ is defined in Assumption \ref{asm-QUAD}.
The derivative of $V(t)$ along \eqref{error-nonlinear-new} gives
\begin{align*}
\dot{V}(t)=&2e^{\mathrm{T}}(\Xi\otimes Q)\left(\mathbf{f}(x)-\phi\left(\tilde{L}_{\sigma(t)}\otimes \Gamma \right)e\right)\\
&\qquad=2\sum_{j=2}^N \xi_j\left(x_1-x_j\right)Q\left(f(x_1)-f(x_j)\right)\\
 &\quad\qquad\qquad-\phi e^{\mathrm{T}}\left[\left(\Xi \tilde{L}_{\sigma(t)}+\tilde{L}_{\sigma(t)}^{\mathrm{T}}\Xi\right)\otimes (Q\Gamma) \right]e.
\end{align*}
By Assumption \ref{asm-QUAD}, one has
$$
\left(x_1-x_j\right)Q\left(f(x_1)-f(x_j)\right)\geq \left(x_1-x_j\right)^{\mathrm{T}}\Sigma\left(x_1-x_j\right).
$$
Moreover, one can find a positive constant $\alpha_{\sigma(t)}$ such that $$\Xi \tilde{L}_{\sigma(t)}+\tilde{L}_{\sigma(t)}^{\mathrm{T}}\Xi\leq \alpha_{\sigma(t)} \Xi.$$
Therefore, one reaches that
\begin{align*}
\dot{V}(t)&\geq 2e^{\mathrm{T}}\left(\Xi\otimes \Sigma \right)e-\phi e^{\mathrm{T}}\left[\alpha_{\sigma(t)}\Xi\otimes (Q\Gamma)\right]e\\
&\geq e^{\mathrm{T}}\left[ \Xi\otimes Q  \left(2\epsilon I-\phi\alpha_{\sigma(t)}\Gamma \right)\right]e,
\end{align*}
where $\epsilon$ is chosen such that $\Sigma>\epsilon Q$.
If $\phi$ is chosen in such a way that for $i=1,\ldots,n$, there holds
$
2\epsilon-\phi\alpha_{\sigma(t)}\gamma_i> 0,
$
i.e.,
$
\phi<\frac{2\epsilon}{\alpha_{\sigma(t)}\gamma_i}$
with $\gamma_i$ being the $i$-th diagonal entry of $\Gamma$, then $\dot{V}(t)> 0$.
By [Theorem 4.3, \cite{Khail}], it is readily obtained that synchronization for nonlinear inter-connected system \eqref{nonlinear-system-dynamics} cannot be realized.
\end{IEEEproof}
\begin{IEEEproof}[Proof of Corollary \ref{nonlinear-corollary}]
Note that since the common part of every reach is empty and
\begin{align*}
\Bigg[I-\sum_{j=1}^{\chi_{\sigma(t)}}  \Big(\gamma_{{\sigma(t)},j}\beta_{{\sigma(t)},j}^{\mathrm{T}}\otimes I_n\Big)\Bigg]\mathbf{F}\left(\sum_{j=1}^{\chi_{\sigma(t)}}  \left(\gamma_{{\sigma(t)},j}\beta_{{\sigma(t)},j}^{\mathrm{T}}\otimes I_n\right)x\right)=0.
\end{align*}
Therefore, $$M_{\sigma(t)}\mathbf{F}(x)=M_{\sigma(t)}\left(\mathbf{F}(x)-\mathbf{F}((I-M_{\sigma(t)})x)\right),$$
which implies that condition \eqref{extension-Lipschitz-condition} holds if Assumption \ref{Lipschitz-assumtion} is satisfied.
The following proof follows straightforward that of Theorem \ref{Tho-Nonlinear}. The remove of  \eqref{extension-Lipschitz-condition} is at the cost that the common part of different reaches is required to be empty in each communication graph.
\end{IEEEproof}
\section{Conclusion\label{sec:conclusion}}
We have investigated the synchronization problem of linear generic systems and Lipschitz-type nonlinear  systems communicating over directed switching topology with {\color{blue}mild} connectivity assumption.
An analysis framework from both algebraic and geometric perspective to deal with the attractiveness of the synchronization manifold has been developed. Specifically, the synchronization problem is transformed into the one of evolution of projection system state onto a set of appropriately selected subspaces.   It turns out that the synchronization of partial-state coupled linear generic systems can be reached if  additional algebraic conditions  are satisfied.  While for  Lipschitz-type  nonlinear systems with positive definite inner coupling matrix, synchronization  can be realized if the coupling strength is strong enough if the subspaces have certain geometric properties.  Two special cases with specific switching scheme and undirected communication graph have also been investigated for linear systems, respectively. Illustrative examples have verified the theoretical findings.
\appendices

\section{Proof of Lemma \ref{semi-simple-zero-eigenvalue}}
\begin{IEEEproof}
Consider the system dynamics
$\dot{x}(t)=-Lx(t),$
with $L$ being a Laplacian matrix of a nonnegatively weighted graph.  Take $T=[\omega_1,\ldots,\omega_{n_1},\ldots,\omega_N]$, where $\omega_1=\frac{1}{\sqrt{N}}\mathbf{1}_{N}$ and $\omega_1,\ldots,\omega_{n_1}$ are chosen to be generalized eigenvectors of eigenvalue zero. Then, the transformation $z=T^{-1}x$ implies
$$
\begin{bmatrix}\dot{z}_1\\\dot{z}_2\end{bmatrix}=-\begin{bmatrix}H_{1}&\\ * &H_{2}\end{bmatrix}\begin{bmatrix}z_1\\z_2\end{bmatrix},
$$
with $H_{1}\in \mathbb{R}^{n_1\times n_1}$  being Jordan block associated with eigenvalue $0$.
Now, suppose $H_{1}$ is not diagonal. Let $x_0$ be chosen such that $z_1=\mathbf{1}_{n_1}$. By $z_1$-dynamics $\dot{z}_1=-H_{1}z_1$, one then has that $\|z_1\|$ grows unbounded with time approaching infinity. This contradicts the result obtained in \cite{Ren2005}, which requires that $\|z_1\|$ is bounded. This completes the proof.
\end{IEEEproof}
\section{Proof of Lemma \ref{kernel-intersect}}
\begin{IEEEproof}
	It is obvious that $\cap_{j=1}^p \mathrm{Ker}(L_j)\supset \mathrm{span}\{\mathbf{1}_N\}$. Now, suppose there exists a nonzero vector $w\notin \mathrm{span}\{\mathbf{1}_N\}$ but $L_jw=0$ for $j=1,\ldots,p$. This in turn implies that $(\sum_{j=1}^p L_j)w=0.$ Since $\cup_{j=1}^p {\color{blue}\mathcal{G}_j}$ contains a directed spanning tree, the kernel of  $\sum_{j=1}^p L_j$ is spanned by $\mathbf{1}_N$ \cite{Ren2005}. Contradiction arises. The first assertion then holds. Note that given any $x\in\mathbb{R}^N$, $L_ix\in\mathrm{Ran}(L_i)$. Then, $(L_1+\ldots+L_p)x\in \mathrm{span}\{\mathrm{Ran}(L_1)\cup\cdots\cup \mathrm{Ran}(L_p)\}$. This implies that $\mathrm{Ran}(\sum_{j=1}^pL_j)\subset \mathrm{span}\{\mathrm{Ran}(L_1)\cup\cdots\cup \mathrm{Ran}(L_p)\}$. It is further concluded that the dimension of $\mathrm{span}\{\mathrm{Ran}(L_1)\cup\cdots\cup \mathrm{Ran}(L_p)\}$ is at least $N-1$. Observe further that $\mathrm{span}\{\mathrm{Ran}(L_1)\cup\cdots\cup \mathrm{Ran}(L_p)\}\cap \mathrm{span}\{\mathbf{1}_N\}=\{{\color{blue}\mathbf{0}} \}$. The second assertion holds.
\end{IEEEproof}

\section{Proof of Lemma \ref{convergence-in-subspace}}
\begin{IEEEproof}
With $\xi=T^{-1}x$ and in view of the invariance of $S_i$ with respect to $H$, one has
\begin{align}
\dot{\xi}=\begin{bmatrix}\tilde{H}_1&&\\&\ddots&\\&&\tilde{H}_p\end{bmatrix}\xi,
\end{align}
where $\mathrm{diag}\{\tilde{H}_1,\ldots,\tilde{H}_p\}=T^{-1}HT$. Then, by the observation that given $x\in S_i,1\leq i\leq p$, $$T^{-1}x=\left[0,\ldots,0,\xi_{\sum_{j=1}^{i-1}n_j+1},\ldots,\xi_{\sum_{j=1}^i n_j},0,\ldots,0\right].$$ The first conclusion follows.

{\color{blue}Notice that $x_i$  and also $x$ can be written as a linear combination of a set of basis vectors $v_{k},k=1,\ldots,n$ such that
$$
x_i=z^i_{1}+\cdots+z^i_{n},
$$
where $z^i_k$ is the projection of $x_i$ onto $v_k$, $k=1,\ldots,n$. Hence, one has
$$
x=\sum_{j=1}^pz_j=\sum_{j=1}^p\sum_{i=1}^n z^j_i,
$$
where $z_j=\sum_{i=1}^n z^j_i.$
Proposition i) in 2) is then straightforward. It is worth pointing out that
\begin{align*}
\|T^{-1}x_i(0)\|^2&\leq \|z_1^i(0)\|^2+\cdots+\|z^{i}_{n}(0)\|^2=\|z_i(0)\|^2\\
&=\|\mathrm{diag}\{0,\ldots,0,1,\ldots,1,0,\ldots,0\}T^{-1}x(0)\|
\end{align*}
with $T=[v_1,\ldots,v_n]$ and the diagonal matrix, denoted by $M_i$ has $n_i$ 1's from the ($\sum_{j=1}^{i-1}n_j+1$)-th to the  ($\sum_{j=1}^{i}n_j$)-th entry. Then, $\Theta_{\mathbb{R}^n}(\psi)=\max_i\sigma_{\max}(M_iT^{-1})/\sigma_{\min}(T^{-1})\psi$. Next, we prove the converse.  Suppose $\|x\| \leq \psi \exp\{\gamma t \}\|x_0\|$, by the observation that $\|M_iT^{-1}x(0)\|\geq\|T^{-1}x_i(0)\|$,
it then follows from similar arguments that $x_i\leq \Phi_{S_i}{\psi}\exp\{\gamma t\}\|x^{UB}_i(0)\|$, where $\Phi_{S_i}(\psi)=\sigma(\max)(M_iT^{\mathrm{T}})/\sigma(\min)(T^{\mathrm{T}})\psi\cdot \max_i \|x_0\|/\|x^{UB}_i(0)\|$. Finally, combining the results in i) and ii) gives immediately iii). This completes the proof.}
\end{IEEEproof}
\section{Proof of Lemma \ref{split-space}}
\begin{IEEEproof}
First, let $\bar{S}_1=S_1\cap S_2$. Then, define $\bar{S}_2$ such that $\bar{S}_2\oplus \bar{S}_1=S_1$. Let $\bar{S}_3$ be given in such a way that $\bar{S}_1\oplus \bar{S}_2\oplus \bar{S}_3=\mathrm{span}\{\cup_{j=1}^{2}S_j\}$. Next, one can  construct $\bar{S}_4=S_3 \cap \bar{S}_1$, $\bar{S}_5=S_3 \cap \bar{S}_2$, and $\bar{S}_6=S_3 \cap \bar{S}_3$.
With the constructed  $\bar{S}_4$, $\bar{S}_5$, and $\bar{S}_6$, the subspaces $\bar{S}_1$, $\bar{S}_2$, and $\bar{S}_3$ are renewed as the subspaces $\bar{S}_1^{*}$, $\bar{S}_2^{*}$, and $\bar{S}_3^{*}$, which are given by $\bar{S}_1=\bar{S}_1^{*}\oplus  \bar{S}_4$, $\bar{S}_2=\bar{S}_2^{*}\oplus  \bar{S}_5$, and $\bar{S}_3=\bar{S}_3^{*}\oplus \bar{S}_6$, respectively. Finally, define $\bar{S}_7$ in the same way as $\bar{S}_3$ such that $\oplus_{j=1}^7 \bar{S}_j=\mathrm{span}\{\cup_{j=1}^3 S_j \}$.

By following the preceding procedures, one can similarly construct $\bar{S}_8, \bar{S}_9,\ldots,\bar{S}_{2^p-1}$. It is worth nothing that $\bar{S}_k$ may be a trivial space, i.e., $\bar{S}_k=\{{\color{blue}\mathbf{0}} \}$ for some $k$.
If $\bar{S}_k$ is trivial, then $\bar{S}_k$ is eliminated. Hence, one finally obtains $1\leq \bar{p}\leq 2^p-1$ subspaces whose sum is a direct sum and equals to $\mathrm{span}\{\cup_{j=1}^p S_j\}$. The latter fact is true by the process of construction of $\bar{S}_i,i=1,\ldots,\bar{p}$. The proof is therefore complete.
\end{IEEEproof}

\section{Proof of Lemma \ref{observability-extension}}
\begin{IEEEproof}
The observability of $(C,A)$ and $(C,A-\Pi C)$ implies that \cite{Antsaklis}
$$
O_A=\begin{bmatrix}C\\CA\\\vdots\\CA^{n-1}\end{bmatrix},\;\;\;O_{A-\Pi C}=\begin{bmatrix}C\\C(A-\Pi C)\\\vdots\\C(A-\Pi C)^{n-1}\end{bmatrix}
$$
are full column rank, respectively. We first show that given nonzero $x\in\mathbb{R}^n$, if $O_Ax\neq 0$ then $O_{A-\Pi C}x\neq 0$. This can be done by induction. Note that if $Cx\neq 0$, then the above conclusion is true. Now, suppose $CA^kx\neq 0, CA^jx=0, \forall 0\leq j<k$. Observe that $C(A-\Pi C)^k$  contains terms that end with $CA^s$  where $0\leq s\leq k$ \footnote{For example, with $k=2$, $C(A-\Pi C)^2=CAA-CA\Pi C-C\Pi CA+C\Pi C\Pi C$.}. If $0\leq s<k$, then $CA^sx=0$ in view of $CA^j=0, \forall 0\leq j<k$. Since $CA^k$ is the only left term of $C(A-\Pi C)^k$ which satisfies $CA^kx\neq 0$, one obtains that $C(A-\Pi C)^kx\neq 0$. The converse that given nonzero $x\in\mathbb{R}^n$, if $O_{A-\Pi C}x\neq 0$ then $O_{A}x\neq 0$, can be proved in the same way. Therefore, it is known that $O_A$ and $O_{A-\Pi C}$ has the same rank. The proof is thus complete.
\end{IEEEproof}
\section{Proof of Lemma \ref{lma-shrink-space}}
\begin{IEEEproof}
 	We first show that eigenvalue $0$ of $\tilde{L}_{\sigma(t_{k_j})}$ is semi-simple. Suppose there exists a vector $v\notin 0$ such that $v\in \mathrm{Eig}(\tilde{L}_{\sigma(t_{k_j})})$ is a generalized eigenvector for some $j$. One can then find a positive integer $n_j>1$ for $\tilde{L}_{\sigma(t_{k_j})}$ such that $\tilde{v}_j=\tilde{L}_{\sigma(t_{k_j})}^{n_j}v\neq 0$ is the eigenvector of $\tilde{L}_{\sigma(t_{k_j})}$ corresponding to eigenvalue 0.  Construct $\bar{v}_j=[0,\tilde{v}^{\mathrm{T}}]^{\mathrm{T}}$. Then, one has
 	$$
 	\Delta L^{n_j+1}_{\sigma(t_{k_j})}\Delta\bar{v}_j=\left[*\tilde{L}^{n_j}_{\sigma(t)}\tilde{v},0\right]^{\mathrm{T}}=\left[*\tilde{L}^{n_j}_{\sigma(t)}\tilde{L}_{\sigma(t_{k_j})}^{n_j}v,0\right]^{\mathrm{T}}=0,
 	$$
 	which in turn implies that
 	$$
 	L^{n_j+1}_{\sigma(t_{k_j})}\left(-\left[0,\tilde{v}_j^{\mathrm{T}}\right]^{\mathrm{T}}\right)=0.
 	$$

 	One can then reach the conclusion that $ \mathrm{Eig}(L_{\sigma(t_{k_j})})$ contains a vector $[0,-\tilde{v}_j^{\mathrm{T}}]^{\mathrm{T}}$ that is a generalized eigenvector associated with eigenvalue 0. This implies that the algebraic multiplicity of eigenvalue $0$ is strictly larger than its geometric multiplicity. This contradicts the result obtained in Lemma \ref{semi-simple-zero-eigenvalue}.
 	
 	Now, suppose
 	$
 	\cap_{j=0}^{m_k-1} \mathrm{Eig}(\tilde{L}_{\sigma(t_{k_j})})\neq \{{\color{blue}\mathbf{0}}\},
 	$
 	which implies that there exists a vector $w\neq 0$ such that $\tilde{L}_{\sigma(t_{k_j})}w=0$ for $j=0,\ldots,m_k-1$. Hence, $\sum_{j=0}^{m_k-1}\tilde{L}_{\sigma(t_{k_j})}w=0.$ Note that $\sum_{j=0}^{m_k-1}\tilde{L}_{\sigma(t_{k_j})}$ is the submatrix of
 	 $$\Delta \sum_{j=0}^{m_k-1}L_{\sigma(t_{k_j})}\Delta=\begin{bmatrix}0&*\\0&\sum_{j=0}^{m_k-1}\tilde{L}_{\sigma(t_{k_j})}\end{bmatrix},$$
 	which indicates that $\sum_{j=0}^{m_k-1}\tilde{L}_{\sigma(t_{k_j})}$ is of full rank. Contradiction arises. The proof is thus complete.
 \end{IEEEproof}

\section{Proof of Lemma \ref{synchronization-error}}
\begin{IEEEproof}
This proof exploits the Frobenius normal form of a Laplacian matrix. Note that $\tilde{{\color{blue}\mathcal{G}}}=\cup_{i=1}^p {\color{blue}\mathcal{G}}_i$ contains a directed spanning tree. Then, $L({\color{blue}\tilde{\mathcal{G}}})$ can equivalently be written into the following Frobenius normal form \cite{Qin2015TNNLS}:
\begin{align*}
L(\tilde{{\color{blue}\mathcal{G}}}) =\begin{bmatrix}L_{11}(\tilde{{\color{blue}\mathcal{G}}})&&\\L_{21}(\tilde{{\color{blue}\mathcal{G}}})&L_{22}(\tilde{{\color{blue}\mathcal{G}}})\\
\vdots&\cdots&\ddots\\
L_{q1}(\tilde{{\color{blue}\mathcal{G}}})&\cdots&\cdots&L_{qq}(\tilde{{\color{blue}\mathcal{G}}})
\end{bmatrix}.
\end{align*} It is worth noting that given ${\color{blue}\mathcal{G}}_1$ to ${\color{blue}\mathcal{G}}_p$, to show  $\cap_{j=1}^p\mathrm{span}(\{x|\delta_{j}=0\})=\mathrm{span}\{\mathbf{1}_N\otimes u,\;u\in\mathbb{R}^n \}$, the sequence of appearance of ${\color{blue}\mathcal{G}}_i$ is irrelevant. Denote  $\tilde{{\color{blue}\mathcal{G}}}^i$ the subgraph of $\tilde{{\color{blue}\mathcal{G}}}$ that corresponds to $L_{ii}(\tilde{{\color{blue}\mathcal{G}}}),i=1,\dots,q$. The following proof resorts to contradiction. We assume that $\delta_i=0,i=1,\ldots,p$ does not imply that $x= \alpha \mathbf{1}_N$ for some $\alpha\in\mathbb{R}$.

To proceed, we start from $\tilde{{\color{blue}\mathcal{G}}}^1$ and fix $x\notin \mathrm{span}\{\mathbf{1}_N \}$ such that $\delta_i=0,i=1,\ldots,p$. Select node $m$ in such a way that $m=\arg\min\{x_i,i\in \tilde{{\color{blue}\mathcal{G}}}^1 \}$ (if there are many choices, select one arbitrarily) and there exists at least one node, say, $j$, such that $x_m<x_j$ for $j\neq m$ and $j \in \tilde{{\color{blue}\mathcal{G}}}^1$ (otherwise, one has $x_i=x_j$ for any $i,j\in\tilde{G}^1$, which completes the first part of the proof).  Since $\tilde{{\color{blue}\mathcal{G}}}^1$ is strongly connected,  one can find a directed path from $m$ to $m$ passing every other node once and only once. The path is denoted by a sequence of edges $(m=j_0,j_1),(j_1,j_2),\ldots,(j_{k-1},j_k=m)$. It is worth pointing out that any link $(j_k,j_{k+1})$ cannot start from one node from the common part of some reach to another node in the exclusive part.
Starting from $m$, there exists a $1\leq s\leq k-1$ such that $x_{j_l}=x_m$ for $l<s$ while $x_m<x_{j_s}$.  Recall that $\delta_i=0$ implies that any two nodes in the same exclusive part of a reach in ${\color{blue}\mathcal{G}}_i$ have identical state. Furthermore, it is known that the state of a node in a common part is a convex combination (with combination coefficient being positive) of those of the nodes in exclusive parts of the corresponding reaches by $4)$ in Lemma \ref{kernel-space}. Then, it is obvious that $x_{j_l}>x_{m}$ for $s<l\leq k$.  Therefore, $x_m<x_m$, which is impossible. Hence, it is obtained that for any two nodes in $\tilde{{\color{blue}\mathcal{G}}}^1$, their states are identical.

Then, it remains to show that the states of any two nodes from $\tilde{{\color{blue}\mathcal{G}}}^i$ and $\tilde{{\color{blue}\mathcal{G}}}^j$ are identical for $i\neq j$ by induction. Given $\tilde{{\color{blue}\mathcal{G}}}^i$ and $\tilde{{\color{blue}\mathcal{G}}}^{j}$ with $j=i+1$, suppose any two nodes in $\tilde{{\color{blue}\mathcal{G}}}^i$ have identical state. Then, consider the augmented graph  $\bar{{\color{blue}\mathcal{G}}}_j$ that consists of $\tilde{{\color{blue}\mathcal{G}}}_j$, a single node $i^*$ that represents $\tilde{{\color{blue}\mathcal{G}}}^i$, and edges from the node $i^*$ to those in $\tilde{{\color{blue}\mathcal{G}}}_j$ if and only if there exists an edge from $\tilde{{\color{blue}\mathcal{G}}}^i$ to any of them. Next, select node $m\in\mathcal{V}(\tilde{{\color{blue}\mathcal{G}}}_j)$ in such a way that $m=\arg\min\{x_i,i\in \tilde{{\color{blue}\mathcal{G}}}^j,x_{i^*} \}$ (If $x_m=x_{i^*}$, then select node $m$ in such a way that $m=\arg\max\{x_i,i\in \tilde{{\color{blue}\mathcal{G}}}^j,x_{i^*} \}$). Actually, $m$ cannot be reached by $i^*$ directly, i.e., $(i^*,m)\notin\mathcal{E}$. Otherwise, there exists a ${\color{blue}\mathcal{G}}_i$ such that $(i^*,m)\in\mathcal{E}({\color{blue}\mathcal{G}}_i)$, which implies that $x_{m}$ will be larger than the smallest state of the nodes in the corresponding exclusive parts of the reaches $m$ belongs to. This contradicts the fact that $x_m\leq x_j$ for $\forall j\neq m,j\in\mathcal{V}(\tilde{{\color{blue}\mathcal{G}}}_i)$. Then there exists at least one node $n\in\mathcal{V}(\tilde{{\color{blue}\mathcal{G}}}_j)$ satisfying $x_m<x_n$. This is true by observing that $x_m<x_{i^*}$ and there exists a node, say, $n$, such that $(i^*,n)\in\mathcal{E}(\tilde{{\color{blue}\mathcal{G}}}_i)$. By following the arguments developed to prove that the states of any two nodes in $\tilde{{\color{blue}\mathcal{G}}}^1$ are the same, it can be verified that $x_m=x_n=x_{i^*}$. This completes the proof.
\end{IEEEproof}



\vspace{-0.6cm}

\end{document}